\UseRawInputEncoding
\documentclass[aps,prr,superscriptaddress,floatfix,twocolumn,longbibliography]{revtex4-2}

\usepackage{amsmath,dcolumn,bm,amsthm,tabularx,subfigure}
\usepackage[colorlinks=true,urlcolor=blue,citecolor=blue,linkcolor=blue]{hyperref}
\usepackage{mathrsfs,times}
\usepackage{bbold,dsfont}
\usepackage{graphicx,graphics,color}
\usepackage{mathtools,nicefrac}
\usepackage{slashed}
\usepackage{amsfonts,amssymb}
\allowdisplaybreaks

\newcommand{\beq}{\begin{equation}}
\newcommand{\eeq}{\end{equation}}
\newcommand{\bea}{\begin{eqnarray}}
\newcommand{\eea}{\end{eqnarray}}
\newcommand{\widebar}{\overline}

\newcommand{\X}{\widetilde{X}}
\newcommand{\Z}{\widetilde{Z}}
\newcommand{\qp}{\mathrm{qp}}
\newcommand{\SvN}{[S_\textrm{vN}]}
\newcommand{\SvNdef}{S_\textrm{vN}}
\newcommand{\trho}{\widetilde{\rho}}
\newcommand{\I}{{I}}
\newcommand{\R}{{R}}

\begin{document}

\title{Many-body localization in the infinite-interaction limit and \\ the discontinuous eigenstate phase transition}

\author{Chun~Chen}
\email[Corresponding author.\\]{chunchen@sjtu.edu.cn}
\affiliation{School of Physics and Astronomy, Key Laboratory of Artificial Structures and Quantum Control (Ministry of Education), Shenyang National Laboratory for Materials Science, Shanghai Jiao Tong University, Shanghai 200240, China}
\affiliation{Max-Planck-Institut f\"{u}r Physik komplexer Systeme, N\"{o}thnitzer Stra{\ss}e 38, 01187 Dresden, Germany}

\author{Yan~Chen}
\email[Corresponding author.\\]{yanchen99@fudan.edu.cn}
\affiliation{Department of Physics and State Key Laboratory of Surface Physics, Fudan University, Shanghai 200433, China}
\affiliation{Collaborative Innovation Center of Advanced Microstructures, Nanjing University, Nanjing 210093, China}

\author{Xiaoqun~Wang}
\email[Corresponding author.\\]{xiaoqunwang@sjtu.edu.cn}
\affiliation{School of Physics and Astronomy, Key Laboratory of Artificial Structures and Quantum Control (Ministry of Education), Shenyang National Laboratory for Materials Science, Shanghai Jiao Tong University, Shanghai 200240, China}
\affiliation{Tsung-Dao Lee Institute, Shanghai Jiao Tong University, Shanghai 200240, China}
\affiliation{Beijing Computational Science Research Center, Beijing 100084, China}

\date{\today}

\begin{abstract}

Can localization persist when interaction grows infinitely stronger than randomness? If so, is it many-body Anderson localization? How about the associated localization transition in the infinite-interaction limit? To tackle these questions, we study many-body localization (MBL) in a spin-chain model mimicking the Rydberg-blockade quantum simulator with both infinite-strength projection and moderate quasiperiodic modulation. Employing exact diagonalization, Krylov-typicality technique, and time-evolving block decimation, we identify evidence for a constrained MBL phase stabilized by a pure quasirandom transverse field. Remarkably, the constrained MBL transition may embody a discontinuous eigenstate phase transition, whose discontinuity nature significantly suppresses the finite-size drifts that plague most numerical studies of conventional MBL transition. Through quantum dynamics, we find that rotating the modulated field from parallel toward perpendicular to the projection axis induces an eigenstate transition between the diagonal and constrained MBL phases. Intriguingly, the entanglement-entropy growth in constrained MBL follows a double-log form, whereas it changes to a power law in approaching the diagonal limit. By unveiling the significance of confined nonlocal effects in integrals of motion of constrained MBL, we show that this newfound insulating state is not a many-body Anderson insulator. Our predictions can be tested in Rydberg experiments.            

\end{abstract}

\maketitle

\noindent \textbf{INTRODUCTION}

\noindent The framework of many-body localization lays its foundation on noninteracting Anderson insulator \cite{Anderson} to address the quest of ergodicity breaking \cite{Basko,Gornyi} and instability toward delocalization and eigenstate thermalization \cite{Deutsch,Srednicki} under weak many-body interactions in low spatial dimensions \cite{Abanin,Huse}.

This short-range weak-interaction picture forms the backbone of conventional MBL. Nonetheless, it also raises a question of whether there can arise many-body non-Anderson localization in circumstances where interaction strengths are not weak but infinitely strong. See \cite{Turner,Horssen,Hickey,Pancotti,SmithMoessner,Brenes,SierantPRA,LiDengWuDasSarma,BarLev} for different considerations on uniform or random-interaction systems.

Phenomenologically, isolated many-body Anderson insulators may be describable by the emergent extensive set of local integrals of motion (LIOMs or $\ell$-bits) \cite{Ros,ImbrieRosScardicchio,Serbyn,HuseLIOM}, at least in one dimension ($1$D) \cite{Imbrie}. Then, is it conceivable that localization persists but owing to restriction or frustration, the LIOM-based picture breaks down? It is known that finite interaction activates more resonance channels for dephasing, so it is expected to suppress localization. In this regard, a better and affirmative route to achieving the unconventional MBL might be associated with the presence of restriction or frustration, for instance, in disordered Rydberg-blockade chains \cite{Bernien,Chen}, where, as a consequence of strong, coherent dipole-dipole van der Waals repulsions, two nearest-neighbouring Rydberg atoms cannot be simultaneously excited. This energy constrained dynamics is modelled by a projection action of infinite strength.

Specifically, would there be a singular boundary separating different phases of MBL due to abrupt distortion rather than a progressive dressing of $\ell$-bits?

MBL phase and MBL transition are two interdependent topics central to ergodicity breaking in statistical mechanics. Recently, there is a debate questioning MBL as a viable state in the thermodynamic limit. The issue stems from the strong finite-size drift of the MBL-thermal phase boundary seen in nearly all numerical scaling analyses of small chains \cite{Suntajs,SuntajsPRB,Panda_2020,SierantThoulessTime,SierantPoly,Abanin2021,Sierant,Mondaini,Luitz,KieferEmmanouilidisPRL,LuitzPRB2020,KieferEmmanouilidisPRB,Sels,Vidmar,Morningstar}. Because the critical disorder strength keeps shifting toward infinity under the increase of system's size, it was inferred that no MBL transition occurs within these models such that MBL might be a finite-size crossover phenomenon that ultimately gives way to the normal process of thermalization.

How about the nature of MBL transition in the presence of infinite interparticle interaction?

Counterintuitively, we find through solving a concrete lattice model that the eigenstate transition between MBL and thermal regimes may contrastingly be discontinuous when (off-diagonal) constrained limit is taken, a feature probably enabled by the infinite interaction that considerably reduces the adverse effects of finite-size drifts at transition points, thereby strengthening the stabilization of unconventional constrained MBL state and, as a byproduct, the robustness of diagonal MBL phase.     

\vbox{} \noindent \textbf{RESULTS}

\noindent \textbf{The minimal model} 

\noindent The aforementioned physics might be visible in disordered and locally constrained chain models \cite{Chen}. The simplest of such category takes the following archetypal form,
\beq
H_\qp=\sum_i \left(g_i\X_i+h_i\Z_i\right),
\label{constrham}
\eeq
where $\X_i,\Z_i$ are projected Pauli matrices, $\X_i\coloneqq P\sigma^x_i P$ and $\Z_i\coloneqq P\sigma^z_i P$. The global operator $P\coloneqq\prod_i(\frac{3+\sigma^z_i+\sigma^z_{i+1}-\sigma^z_i\sigma^z_{i+1}}{4})$ prohibits motifs of $\downarrow\downarrow$-configuration over any adjacent sites, hence rendering the Hilbert space of model (\ref{constrham}) locally constrained.

In Ref.~\cite{Chen}, we showed that a random version of model (\ref{constrham}) by quenched disorder exhibits tentative signatures of a constrained MBL (cMBL) phase; nevertheless, as being in proximity to the nearby transition, the Griffiths effect therein proliferates, which impedes an identification and thus a direct investigation of this unconventional nonergodic state of matter. In current work, we improve our prior construction by conceiving an experiment-pertinent quasiperiodic constrained model with open and periodic boundary conditions (BCs), i.e., choosing \cite{Iyer,Khemani,Lee,Nag,Dutta,ZhangYao,Agrawal,Mace,Singh,Duthie,SierantPRB,Aramthottil,Thomson,Znidaric2018,Szabo,Doggen,Sierant2022}
\begin{equation}
g_i=g_x+W_x\cos(2\pi\xi i+\phi_x),\ \ h_i=W_z\cos(2\pi\xi i+\phi_z),
\label{factor_gi_hi}
\end{equation}
where the wavenumber $\xi=\sqrt{2}$ is irrational, $i=1,\ldots,L$, and $\phi_x,\phi_z\in[-\pi,\pi)$ are different sample-dependent random overall phase shifts.

Since Hamiltonian (\ref{constrham}) is real, time-reversal symmetry ${\sf T}\coloneqq K$ is preserved, giving rise to the Gaussian orthogonal ensemble (GOE) in phases obeying the eigenstate thermalization hypothesis (ETH) \cite{Khemani2}. Additionally, when $W_z=0$ there is a particle-hole symmetry ${\sf P}\coloneqq\prod_i\sigma^z_i$ that anticommutes with $H_\qp$. To our knowledge, no discrete Abelian symmetry is present in Hamiltonian (\ref{constrham}), so the possibility of a localization-protected spontaneous symmetry breaking \cite{HusePRB} is excluded.     

To manifest the fundamental interplay between finite tunable randomness and infinite interparticle interaction, we introduce hardcore boson operators $b^\dagger,b$ on each site to describe the local pseudospin-$\frac{1}{2}$ system that emulates the Rydberg lattice gas with ground state $|g\rangle=|\!\uparrow\rangle$ and Rydberg excitation state $|r\rangle=|\!\downarrow\rangle$. In terms of hardcore bosons, the Pauli spin matrices can be couched as follows,
\begin{gather}
b^\dagger+b=|r\rangle\langle g|+|g\rangle\langle r|=|\!\downarrow\rangle\langle\uparrow\!|+|\!\uparrow\rangle\langle\downarrow\!|=\sigma^x, \label{boson_sigmax} \\[0.5em]
b^\dagger b=n=|r\rangle\langle r|=|\!\downarrow\rangle\langle\downarrow\!|=\frac{(1-\sigma^z)}{2}, \label{boson_sigmaz}
\end{gather}
where $n=0,1$ is the local occupation number of boson. Armed with the above expressions, Hamiltonian (\ref{constrham}) can then be exactly mapped onto an array of neutral atoms in the Rydberg blockade regime,
\begin{align}
H_{\textrm{qp}}&=H_x+H_z+H_V, \label{constrhamboson} \\[0.5em]
H_x&=\sum_ig_i(b^\dagger_i+b_i), \\
H_z&=\sum_ih_i(1-2n_i), \\
H_V&=\sum_iV_1 n_i n_{i+1},\ \ \ V_1=\infty. \label{vdW}
\end{align}
Here $g_i,h_i$ are proportional to onsite Rabi frequency and frequency detuning, respectively; the repulsive van der Waals interaction in Eq.~(\ref{vdW}) is truncated to retain only the nearest-neighbour interaction whose strength $V_1$ is lifted to infinity, producing a blockade radius of $a<R_b<2a$. Clearly, $H_x$ breaks system's particle-number conservation, so the total energy is the only conserved quantity of the model.

Instructively, using spin operators in Eqs.~(\ref{boson_sigmax}) and (\ref{boson_sigmaz}), Hamiltonian (\ref{constrhamboson}) can also be recast into the standard mixed-field Ising model, for which Imbrie \cite{Imbrie} proved in a mathematically quasiexact way the existence of many-body-generalized Anderson insulator under conditions of limited level attraction, weak interaction strengths, and sufficiently strong disorders. In this regard, although we focus on the infinitely interacting version of such a particular model, the gained results bear the originality and significance to stimulate the research about many-body non-Anderson localization. Further, without $H_V$, $H_0=H_x+H_z$ is a free Hamiltonian describing decoupled spins, each undergoing an independent Larmor precession about the local random fields.

Therefore, the constrained Rydberg atomic chain we consider consists of two pieces: a randomized but noninteracting term $H_0$ and a nearest-neighbour density-density interacting term $H_V$ featured by an infinite repulsion.

Such a compact form with a single \lq\lq spin-like'' sector and the reduction of onsite Hilbert-space dimension from the usual value of $2$ to the golden ratio $\phi=1.618\ldots$ prompt us to regard the bare bones model (\ref{constrhamboson}) [or Eq.~(\ref{constrham})] as the fundamental building block for studying the more generic constrained quantum systems, such as the $t$-$J$ model.

\begin{figure*}[tb]
\centering
\includegraphics[width=0.9\linewidth]{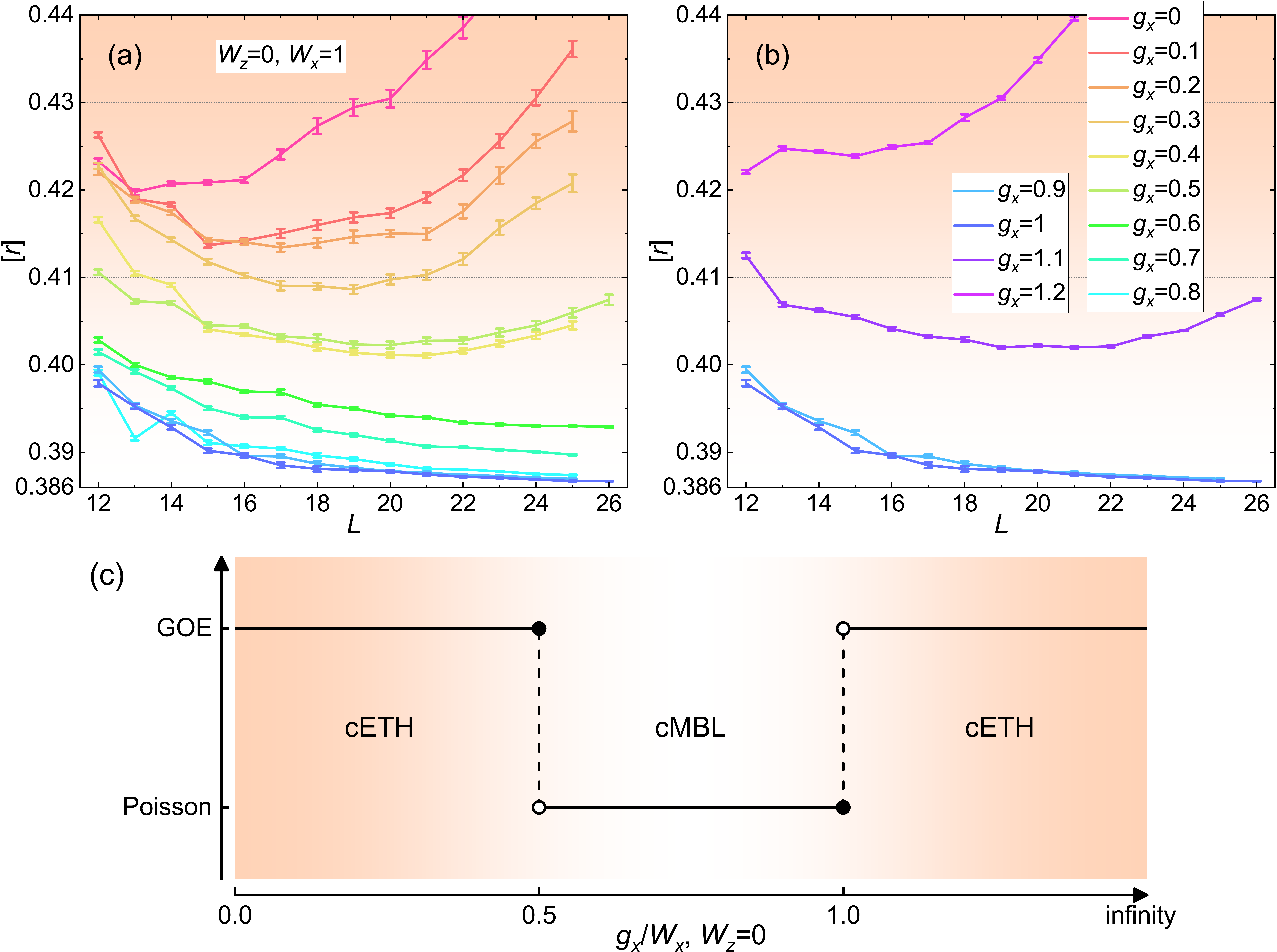}
\caption{\label{fig:fig1a} Spectral diagnostics of the cMBL-cETH transition based on the scaling analysis of $[r]$ using OBCs. The maximal chain length is pushed to $L=26$. Under constrained limit $W_z=0$, the model's $[r]$-value forms a systematic trend in approaching $r_{\textrm{Poi}}$ when $\frac{g_x}{W_x}\in(0.5,1]$, demonstrating the realization of cMBL. Panels (a) and (b) show the proposed respective discontinuity for the left and right boundary of the transition. Panel (c) schematizes the ideal phase diagram along the (off-diagonal) constrained line in the thermodynamic limit where, as denoted by black dots, the chain at the left discontinuous transition point $g_x=0.5$ is thermalized whereas it remains fully localized at the right discontinuous transition point $g_x=1$. This cMBL phase survives to finite $\frac{W_z}{W_x}\approx0.5$, hence forming a dome separated from both the constrained thermal phase at leading $g_x$ and the dMBL state at dominant $W_z$.} 
\end{figure*} 

Moreover, in light of the following commutation relations,
\begin{equation}
[H_x,H_V]\neq0,\ \ \ [H_z,H_V]=0,
\end{equation}
the constrained Hamiltonian (\ref{constrhamboson}) may accommodate two distinct physical extremes. (i) When $|W_z|\gg|g_x|,|W_x|$, the system approaches the diagonal limit during which the role of infinite interaction is effectively minimized and the resulting diagonal MBL (dMBL) state represents a variant of many-body Anderson insulator with enhanced robustness \cite{ChenChenLR}. To be pedantic, throughout this paper, we define the diagonal limit, an analog of Anderson limit, as specified by $h_i\neq0$ and $g_i=0$ in (\ref{constrham}); while for dMBL, the analog of many-body Anderson localization, $|g_i|$, although perturbatively smaller than $|h_i|$, is not identically zero. (ii) In comparison, once $|g_x|,|W_x|\gg|W_z|$, the system enters the off-diagonal constrained limit---the true \lq\lq infinite-interaction limit'' quoted in the paper's title---where mutual impacts from modest randomness and infinite interaction are contrastingly maximized. Particularly, their constructive interplay gives rise to the sought infinite-interaction-facilitated MBL state which is different from the \lq\lq infinite-randomness-controlled'' many-body Anderson localization stabilized in the opposite limit of weak interaction. Na\"ively, no apparent duality would directly link these two.

It is worth stressing that the kinetic constraint was realized in Rydberg-blockade chain \cite{Bernien} and the quasiperiodic modulation played a vital role in experiments \cite{SchreiberBloch,BordiaBloch,LukinGreiner} to achieve the first signature of MBL in unconstrained systems. Accordingly, the actual value of model (\ref{constrham}) resides right in its high experimental relevance. 

Throughout this paper, $W_x=1$ sets the energy scale, i.e., the system is quasirandom at least along $x$ direction.

\vbox{} \noindent \textbf{Discontinuous cMBL-cETH transition: Spectral analyses}

\noindent The configuration averaged level-spacing ratio $[r]$ is the unique single-value quantity routinely adopted to characterize the dynamical states of matter. One defining feature of the robust localization is the vanishing repulsion between contiguous gaps and the resulting Poisson distribution of
\beq
r_n\coloneqq\frac{\min\{\delta_n,\delta_{n-1}\}}{\max\{\delta_n,\delta_{n-1}\}}
\eeq
with mean $[r]=r_{\textrm{Poi}}\approx0.386$ where $\delta_n\coloneqq E_n-E_{n-1}$ assuming $\{E_n\}$ an ascending list of eigenvalues \cite{Oganesyan,Atas,Mondaini,Luitz,Giraud}.

Figures~\ref{fig:fig1a}(a),(b) show the finite-size evolutions of $[r]$ as a function of $g_x$ along the $W_z=0$ axis. Via optimization of the ED algorithm targeting only the eigenvalues, we obtain the full eigenspectra of the chain for $1000$ independent quasirandom samples up to system size $L=26$ and the corresponding Hilbert-space dimension for such a single sample is $317811$ under OBCs. (Parenthetically, the maximal chain length examined by a similar work \cite{Sierant} is also $L=26$, as they used PBCs, the corresponding Hilbert-space dimension increases to $271443$.) Within $0.5<g_x\leqslant1$, we find that $[r]$ steadily converges to $r_{\textrm{Poi}}$ under the successive increase of $L$, verifying the stabilization of a desired cMBL phase.

\begin{figure}[tb]
\centering
\includegraphics[width=1\linewidth]{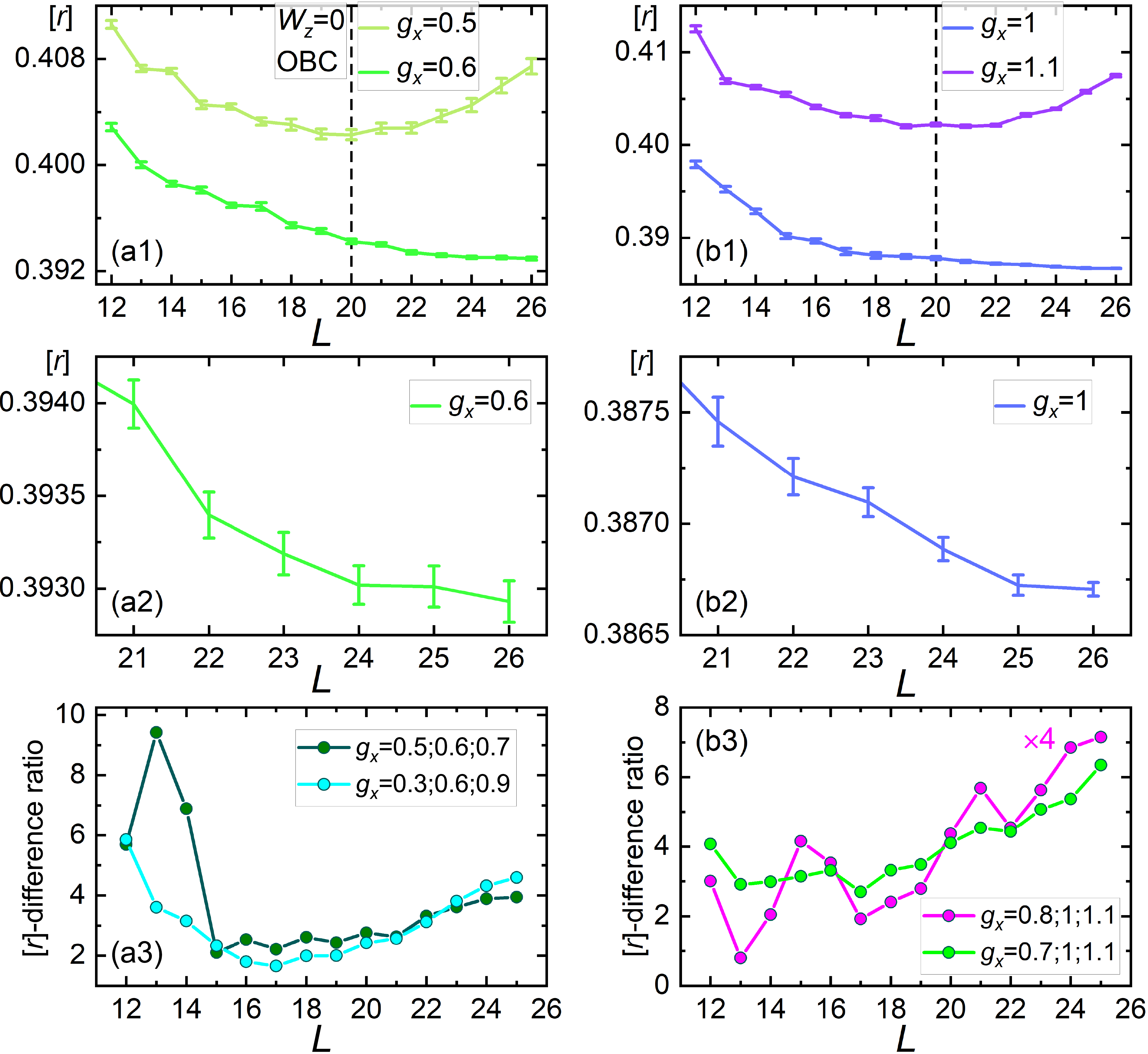}
\caption{\label{fig:fig1b} $[r]$ data at the left [column (a)] and right [column (b)] transition points. Dashed lines in (a1),(b1) mark the lengths exceeding which the $[r]$ values for $g_x=0.5,1.1$ begin to grow. By contrast, (a2),(b2) show the monotonic decrease of $[r]$ at $g_x=0.6,1$ toward $r_{\textrm{Poi}}$. All data are taken from Fig.~\ref{fig:fig1a} with $W_z=0$. (a3),(b3) plot the $[r]$-difference ratios near the two transition points to assess the sharpness of the discontinuity for the underlying transition.} 
\end{figure}

The peculiarity of the quasiperiodic Rydberg chain may be rooted in the discontinuous eigenstate phase transition between cMBL and its nearby constrained ETH (cETH) phase. To pin down the potential discontinuity, we perform a finite-size scaling analysis of $[r]$ by pushing the chain length to $L=26$ and simultaneously selecting a compatibly high resolution $0.1W_x$ when tuning $g_x$. Figure~\ref{fig:fig1a}(a) targets the left discontinuous transition point, from which one observes that for $0\leqslant g_x\leqslant0.5$, there exists a turning point of $L$ beyond which $[r]$ starts to increase continually toward GOE. More precisely, we find the following correspondence between $g_x$ and its turning position of the length:
\beq
\begin{bmatrix}
g_x \\
L
\end{bmatrix}=\begin{bmatrix}
0 \\
13
\end{bmatrix};\begin{bmatrix}
0.1 \\
15
\end{bmatrix};\begin{bmatrix}
0.2 \\
17
\end{bmatrix};\begin{bmatrix}
0.3 \\
19
\end{bmatrix};\begin{bmatrix}
0.4 \\
21
\end{bmatrix};\begin{bmatrix}
0.5 \\
20
\end{bmatrix},
\eeq
namely, the turning point of $L$ increases along with $g_x$ up to $g_x=0.5$. Surprisingly, this trend terminates abruptly once $g_x>0.5$. For instance, even after moving upward to $L=26$, we find no signature of such a turning length for $g_x=0.6$ and its $[r]$-value keeps rolling down toward $r_\textrm{Poi}$. This continual decrease of $[r]$ becomes more transparent for $g_x=0.7,0.8,0.9,1$, which suggests the realization of cMBL within $g_x\in(0.5,1]$. Therefore, the sharp distinction of the two trends induced by an incremental change of $g_x$ indicates that the assumed discontinuity of the transition occurs at $g_x=0.5$.

\begin{figure}[tb]
\centering
\includegraphics[width=1\linewidth]{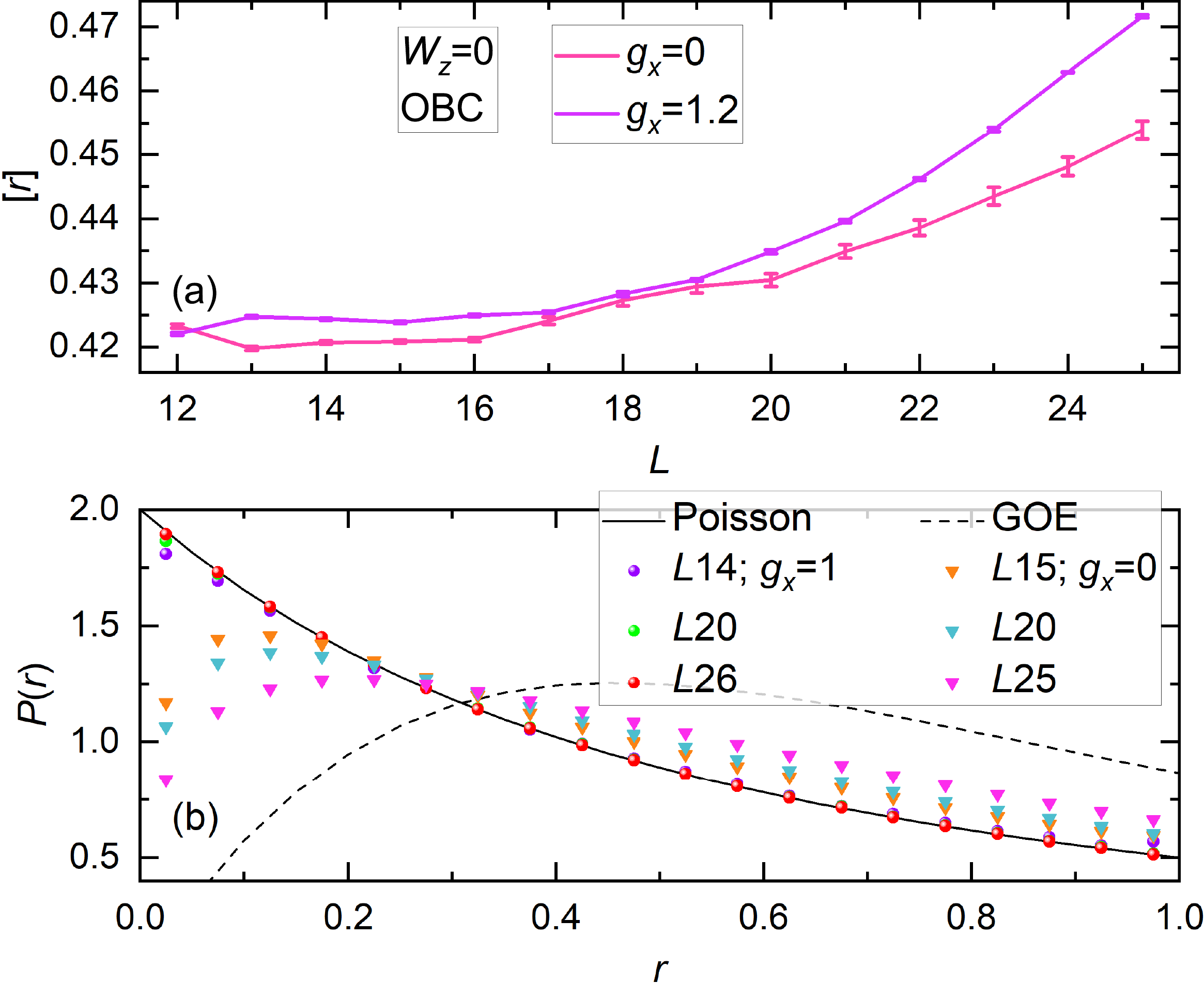}
\caption{\label{fig:fig1c} (a): $[r]$-values of the chain for the two parameter sets $(g_x=0,W_z=0)$ and $(g_x=1.2,W_z=0)$. Both grow promptly toward the value of GOE with $L$, suggesting the system at both parametric points obeys ETH in the thermodynamic limit. (b): The probability distribution of the level-spacing ratio $P(r)$ in cMBL (ensured by $g_x=1,W_z=0$), which closely traces the prediction of Poisson statistics [solid line, $P_\textrm{Poi.}(r)=2/(1+r)^2$] under the increase of $L$, indicating the realization of a full MBL. By selecting $g_x=0,W_z=0$, the $P(r)$ distribution switches to follow the Wigner surmise [dashed line, $P_{\textrm{GOE}}(r)=(27/4)(r+r^2)/(1+r+r^2)^{5/2}$], signalling the thermalization sets in.} 
\end{figure}

The discontinuity between the two contrasting trends showcases more vigorously at the right discontinuous transition point. As presented by Fig.~\ref{fig:fig1a}(b), although working on the small chain of $L=26$, the $[r]$-value for $g_x=1$ is already extremely close to the ideal value of the Poissonian distribution; while, in a striking comparison, the $[r]$-value for $g_x=1.2$ appears to shoot up toward GOE at the very similar length scale. See also Figs.~\ref{fig:fig1c}(a),(b). Crucially, the equality of $[r]$ at $g_x=0.9,1$ and the finite jump of $[r]$ at $g_x=1.1$ comprise a vivid definition of the discontinuity.

Based on the insights gained from extrapolating the scaling analysis of $[r]$, we draw in Fig.~\ref{fig:fig1a}(c) the schematic phase diagram of the quasirandom Rydberg chain in the thermodynamic limit. Here we exclusively focus on the constrained limit by fixing $W_z=0$. After taking random averages, the sign of $W_x$ makes no difference. Together with the unitary transformation rotated by $\sigma^z_i$, the sign of $g_x$ does not matter either. One can thus take $W_x=1$ and consider $\frac{g_x}{W_x}\in[0,+\infty)$ without loss of generality. There are several features about the phase diagram. (i) There only exist two eigenstate phases along the constrained line $W_z=0$, the cMBL phase and the cETH phase. (ii) The cMBL phase occupies a finite interval $g_x\in(0.5,1]$ and the cMBL-cETH transition is likely discontinuous whose transition points locate at the phase boundaries $g_x=\frac{1}{2}$ and $g_x=1$. (iii) The system appears thermalized at $g_x=\frac{1}{2}$; while on $g_x=1$, the chain maintains its full localization character. (iv) At the special point $g_x=0$, where the relative randomness strength is infinite $W_x/g_x=\infty$, the chain is well within the cETH phase.

The robustness of cMBL implies the existence of dMBL in the thermodynamic limit, because adding another source of randomness can only enhance localization. In this sense, the constraint-induced delocalization reported by Ref.~\cite{Sierant} does not lead to contradictions but rather highlights the importance of cMBL as the state to foster dMBL. It also becomes clear about the necessity to include both terms of $g_x$ and $W_x$ on an equal footing to achieve the stabilization of localization in a general constrained setting. Nonetheless, this type of randomness encapsulated by the term $g_i$ is not the topic of \cite{Sierant}.

{\it Transition points refined.}---We examine the postulated discontinuity of $[r]$ at the transition points a bit further in Fig.~\ref{fig:fig1b}, where by zooming in the left [panel (a1)] and right [panel (b1)] transition zones, we highlight the opposite scaling trends of $[r]$ for the two $g_x$'s that are close in magnitude. In accordance with the emergent integrability of cMBL, Figs.~\ref{fig:fig1b}(a2),(b2) show up to $L=26$ the decrease and the convergence of $[r]$ onto the Poisson value at $g_x=0.6,1$. The degree of discontinuity of the transition may be quantified in terms of the $[r]$-difference ratio defined by comparing the $[r]$-values at three adjacent $g_x$'s, viz., with a particular $L$, for ordered $g_x=g_1;g_2;g_3$,
\begin{align}
[r]&\mbox{-difference ratio} \nonumber \\
&\coloneqq\frac{\textrm{max}\!\left\{\big|[r](g_1)-[r](g_2)\big|,\big|[r](g_2)-[r](g_3)\big|\right\}}{\textrm{min}\!\left\{\big|[r](g_1)-[r](g_2)\big|,\big|[r](g_2)-[r](g_3)\big|\right\}}.
\end{align}
For continuous transitions, the $[r]$-difference ratio is on the order of $1$; whereas, if discontinuity arises, then it is predicted to diverge right at the transition once the thermodynamic limit is taken. By choosing four $g_x$-tuples involving the two transition points, we illustrate via Figs.~\ref{fig:fig1b}(a3),(b3) the growth of $[r]$-difference ratio above unity under the increase of $L$. Within the system sizes we probe, the degree of discontinuity on the right transition point appears stronger than that of the left one. This is attributed to the different eigenstate phases realized at these two transition points.

\begin{figure}[b]
\centering
\includegraphics[width=1\linewidth]{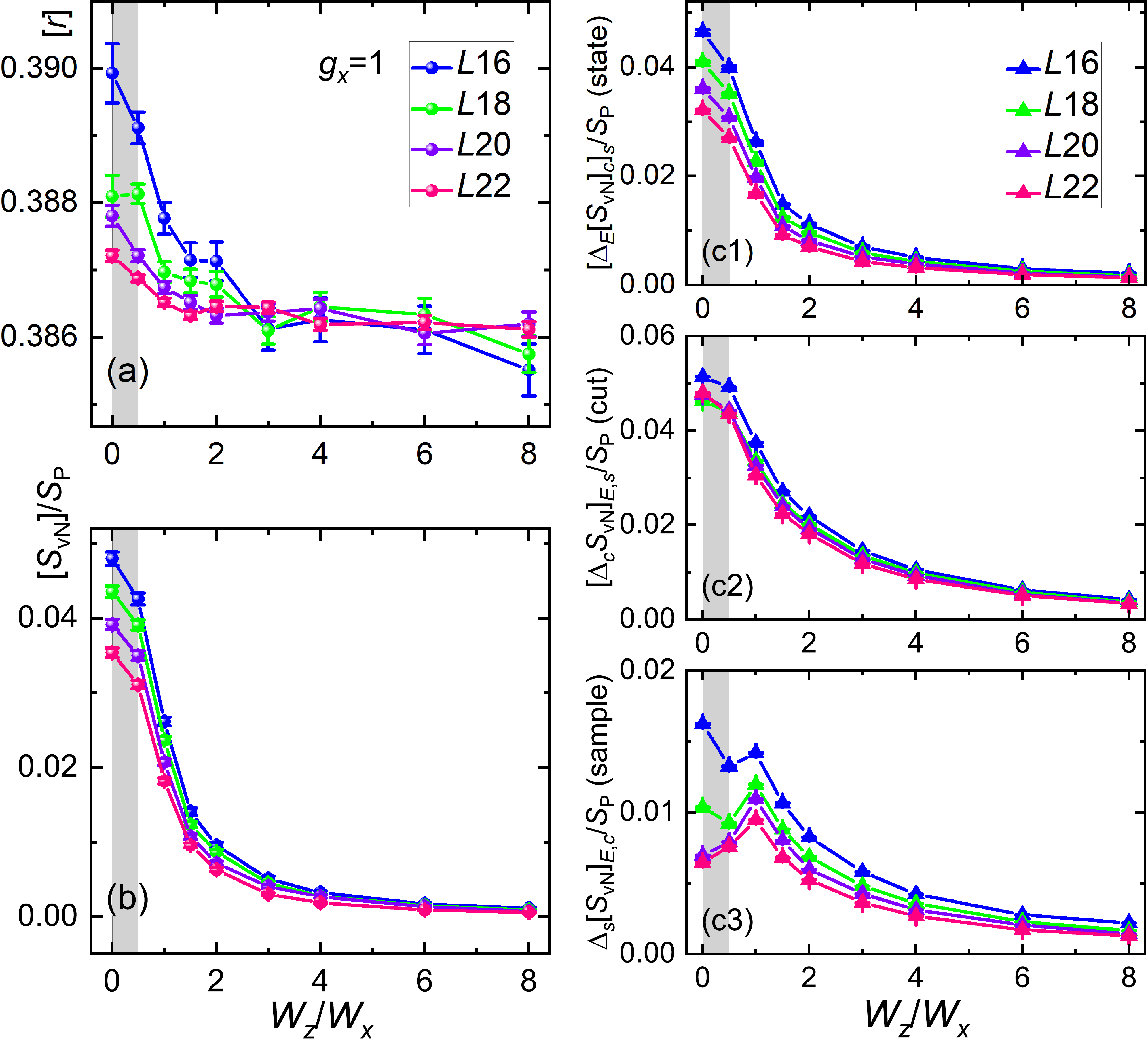}
\caption{\label{fig:fig1d} Static diagnostics of the continuous cMBL-dMBL transition using the variance of $S_\textrm{vN}$ under OBCs. (a),(b): Finite-size scaling analyses show that at fixed $\frac{g_x}{W_x}=1$, $[r]$ and $\SvN/S_\textrm{P}$ stay around $r_{\textrm{Poi}}$ and $0$ under the adjustment of $W_z$ from cMBL toward dMBL. No discontinuity is observed. (c1): The intrasample deviation of the entanglement entropy across the entirety of the eigenspectrum, $[\Delta_E[S_{\textrm{vN}}]_c]_s/S_\textrm{P}$. The accompanying cut-to-cut and sample-to-sample parsings of this entropy deviation are plotted by (c2) and (c3), respectively. The peak forming at $\frac{W_z}{W_x}=1$ in (c3) points toward a transition between cMBL (shaded) and dMBL.} 
\end{figure}

{\it Purely random point.}---On phase diagram Fig.~\ref{fig:fig1a}(c), the point $(\frac{g_x}{W_x}=0,\frac{W_z}{W_x}=0)$ bears multiple physical meanings. (i) It represents the purely random limit. Fig.~\ref{fig:fig1c}(a) evidences that in the presence of infinite interaction, the purely random Rydberg chain likely reaches full thermalization under the constrained limit with no signature of localization. This contrasts to the full localization at $(\frac{g_x}{W_x}=1,\frac{W_z}{W_x}=0)$ as is displayed by Fig.~\ref{fig:fig1c}(b). (ii) An implication of (i) pertains to the probable connection between the existence of cMBL and the discontinuity of the cMBL-cETH transition. This is because if assume the cMBL transition is continuous, then the inevitable finite-size drifts would dominate and consequently cast cMBL into doubt. (iii) When $W_z=0$, by implementing the unitary symmetry involving $\sigma^z_i$ to alter the sign of $g_i$ combined with an energy-scale redefinition, one can exactly map $(\frac{g_x}{W_x}=0,\frac{W_z}{W_x}=0)$ onto $(\frac{g_x}{W_x}=1,\frac{W_z}{W_x}=0)$; however, this mapping is valid if the randomness is quench disorder and obeys the uniform box distribution \cite{Chen}. For quasiperiodic randomness, such a formal equivalence breaks down, but our numerical data hint that $(\frac{g_x}{W_x}=0,\frac{W_z}{W_x}=0)$ might still be equivalent to a point infinitesimally close to but different from $(\frac{g_x}{W_x}=1,\frac{W_z}{W_x}=0)$. Symbolically, it reads $(\frac{g_x}{W_x}=1+0^+,\frac{W_z}{W_x}=0)$. This reasoning implies that the two disjoint cETH regions on the left and right side of the cMBL phase are physically equivalent, and the robustness of cMBL may thus be cemented by the significant degree of discontinuity of the transition at $g_x=1$. In parallel, one can hypothesize that the phase diagram for the quench disordered Rydberg chain is identical to Fig.~\ref{fig:fig1a}(c) except that the black dot at $g_x=1$ moves from Poisson to GOE. This conjecture overlooks the Griffiths rare-region effects and the ensuing avalanche-driven delocalization \cite{DeRoeck} in disordered models which can be detrimental to cMBL phase and transition alike.                         

\vbox{} \noindent \textbf{Continuous cMBL-dMBL transition: Entanglement variances}

\noindent The bipartite entanglement entropy $[S_{\textrm{vN}}]$ is another useful proxy for analyzing MBL as is $[r]$ \cite{Mondaini,Luitz,Yu}. For each eigenstate $|\psi_n\rangle$, the half-chain von Neumann entropy is defined by $\SvNdef\coloneqq-{\sf Tr}_R\left[\rho_R\log_2\rho_R\right]$ where $\rho_R\coloneqq{\sf Tr}_L[|\psi_n\rangle\langle\psi_n|]$ is the reduced density matrix of the right half chain. Using ED, we compute $\SvNdef$ for the entire eigenspectrum of a given sample and obtain $\SvN$ after averaging over all available eigenstates for more than $1000$ independent quasirandom realizations.

Figures~\ref{fig:fig1d}(a),(b) show the respective evolutions of $[r]$ and $\SvN/S_\textrm{P}$ as a function of $W_z$ along the $g_x=1$ axis. Upon increasing $L$, both $[r]$ and $\SvN/S_\textrm{P}$ converge to a flattening curve centring around $r_\textrm{Poi}$ and $0$, respectively, demonstrating the system retains full localization as $W_z$ varies from cMBL to dMBL. Because no discontinuity is found in $[r]$ and $\SvN/S_\textrm{P}$, the change of the phase structure is expected to be continuous. Here $\SvN$ is normalized with respect to the Page value of the thermal entropy \cite{Page} whose estimate suitable for constrained circumstance equals $S_{\textrm{P}}\approx\log_2(F_{L/2+2})-1/(2\ln2)$ where the half-chain Hilbert-space dimension equals the Fibonacci number $F_{L/2+2}$ \cite{Chen}.

Differing in entanglement patterns, continuous transition between unconstrained MBL and ETH phases can be probed via the standard deviation of $S_{\textrm{vN}}$. As first demonstrated by Refs.~\cite{Kjall,KhemaniPRX}, the explicit dependence of $S_{\textrm{vN}}$ on eigenstate wavefunction (\lq\lq $E$''), randomized sample (\lq\lq $s$''), and partition cut (\lq\lq $c$'') brings about three measures for the quantity. In the following, we adopt this strategy for the continuous cMBL-dMBL transition. We borrow the convention of Ref.~\cite{Chen}, viz., using $[\ldots]_{E/s/c}$ to denote the respective averages over the eigenspectrum entirety, all samples, and all cuts with the unspecified subscripts holding fixed. Similar definitions carry over to the standard deviations $\Delta_{E/s/c}(\ldots).$

In accord with the continual changes of $[r]$ and $[S_{\textrm{vN}}]/S_\textrm{P}$ in Figs.~\ref{fig:fig1d}(a),(b), Fig.~\ref{fig:fig1d}(c1) illustrates the subvolume scaling law of $[\Delta_E[S_{\textrm{vN}}]_c]_s/S_\textrm{P}$, the state-to-state intrasample deviation of $S_{\textrm{vN}}$, upon raising $L$ and its overall smooth lineshape as a function of $W_z$. Likewise, the cut-to-cut entanglement-entropy deviation as given by Fig.~\ref{fig:fig1d}(c2) exhibits the qualitatively consistent tendency for the same evolution between the two MBLs.

By comparison, albeit being less prominent for quasiperiodic arrangements, the sample-to-sample entanglement-deviation curve presented by Fig.~\ref{fig:fig1d}(c3) gives the indication of an emergent peak around $\frac{W_z}{W_x}=1$, hinting that the rise of $W_z$ at fixed $g_x=1$ may drive a continuous transition from cMBL toward dMBL. Intuitively, although both MBLs are dominantly constituted by area-law entangled eigenstates, on finite-length chains, the rates of how they approach the area scaling law may differ in magnitude and form, which potentially allows for the entropy variance across different samples near the phase boundary. Nonetheless, in view of the fact that the absolute value of the deviation is not pronounced and the shape of the curve keeps flattening, it is possible that cMBL and dMBL might vaguely be distinguishable from pure static diagnostics.     

\vbox{} \noindent \textbf{Eigenstate transition from entanglement growth}

\noindent Alternatively, the qualitative difference between cMBL and dMBL can be demonstrated from the angle of real-time evolution of entanglement. Notably, we find an eigenstate transition between these two dynamical regimes in the numerical quantum quench experiments.

We use two quantities, the bipartite entanglement entropy and the quantum Fisher information (QFI). The initial state is randomly selected from the complete basis of nonentangled product states of $\sigma^z_i$-spins that respects the local constraint. For each $L$, we generate more than $1000$ random pairs of $(\phi_x,\phi_z)$ for the Hamiltonian, and for each quasiperiodic arrangement, we let the chain evolve and calculate $\SvNdef$, QFI by ED and TEBD \cite{Vidal} before averaging (see methods section).

\begin{figure}[tb]
\centering
\includegraphics[width=1\linewidth]{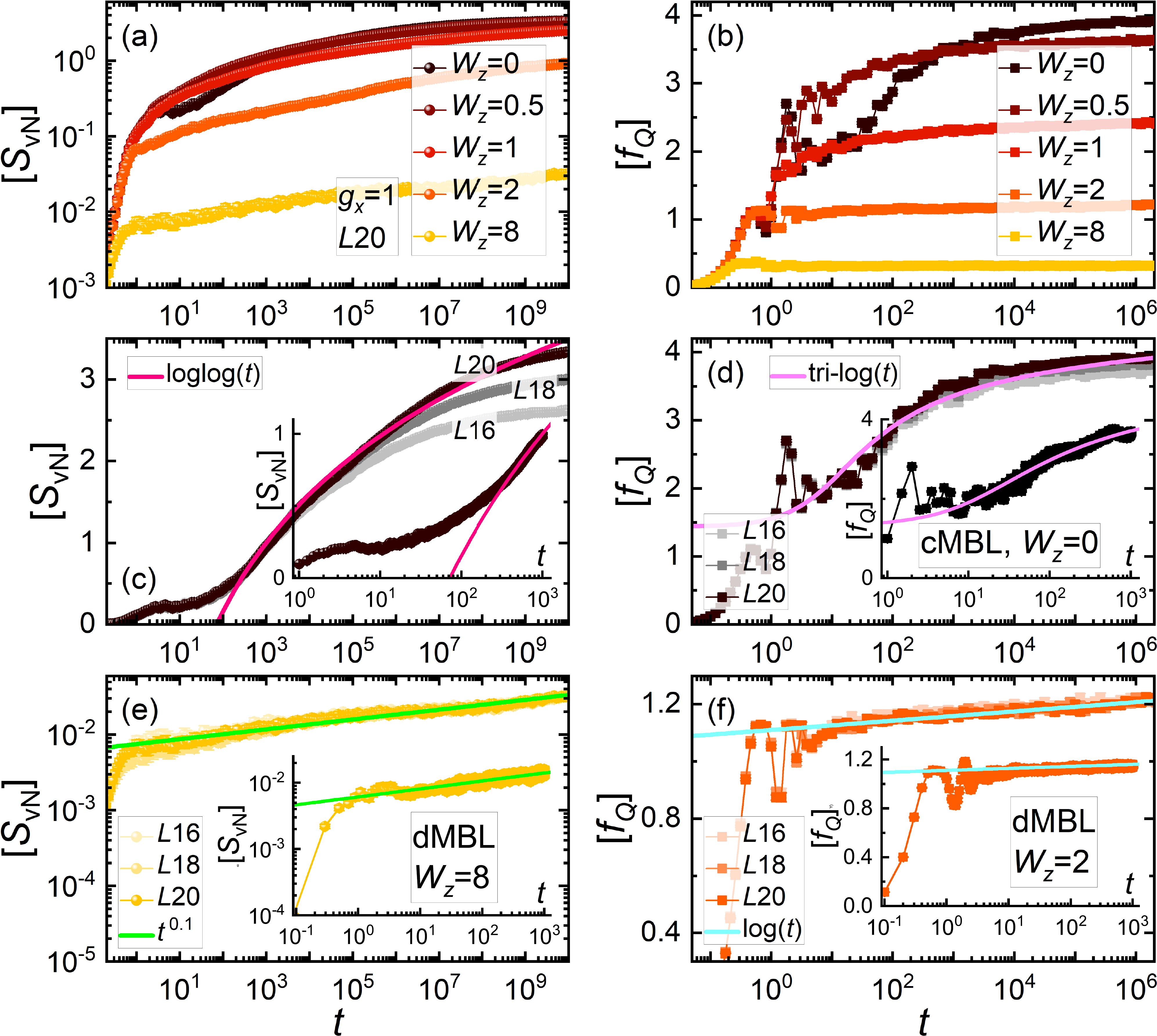}
\caption{\label{fig:fig2a} cMBL-dMBL transition in dynamics with OBCs and fixed $g_x=1$. The maximum chain length in ED is $L=20$. The top row summarizes functional changes of the growth of $[S_{\textrm{vN}}]$ and $[f_Q]$ as a function of $W_z$. Fits in the middle row suggest that for cMBL at $W_z=0$, the entanglement (QFI) growth follows a double (triple) logarithmic form. The bottom row targets the dynamics of dMBL at large $W_z$: consistent with the logarithmic rise of $[f_Q]$, $[S_{\textrm{vN}}]$ grows as a power law of $t$ in dMBL. The four insets of (c)-(f) present the corresponding TEBD results of $L=28$.}
\end{figure}

Figure~\ref{fig:fig2a}(a) compiles time evolutions of $\SvN$ along the cut $g_x=1$ with ascending $W_z$ at $L=20$ in a log-log format. The salient feature there is the qualitative functional change in the time-evolution profiles. This eigenstate transition is elaborated by Figs.~\ref{fig:fig2a}(c) and (e) where we focus on the entanglement growth deep inside cMBL and dMBL, respectively. For concreteness, after a transient period $t\lesssim1$ of the initial development, $\SvN$ in dMBL grows steadily as a power law of $t$ [with an exponent $(\approx0.1)$] within the next prolonged window (up to $t\approx10^{14}$ at $L=20$) but its saturated value is far less than the thermal entropy $S_T\approx\log_2(F_{2+L/2})-1/(2\ln2)-0.06$ \cite{Chen}. In stark comparison, the growth of $\SvN$ in cMBL as displayed by Fig.~\ref{fig:fig2a}(c) follows a different functional form: within $10^2\lesssim t\lesssim10^9$ at $L=20$, the double-log function fits the entropy data reasonably well. Moreover, the equilibrated $\SvN$ reaches a subthermal value in cMBL and obeys a volume scaling law.

\begin{figure}[b]
\centering
\includegraphics[width=1\linewidth]{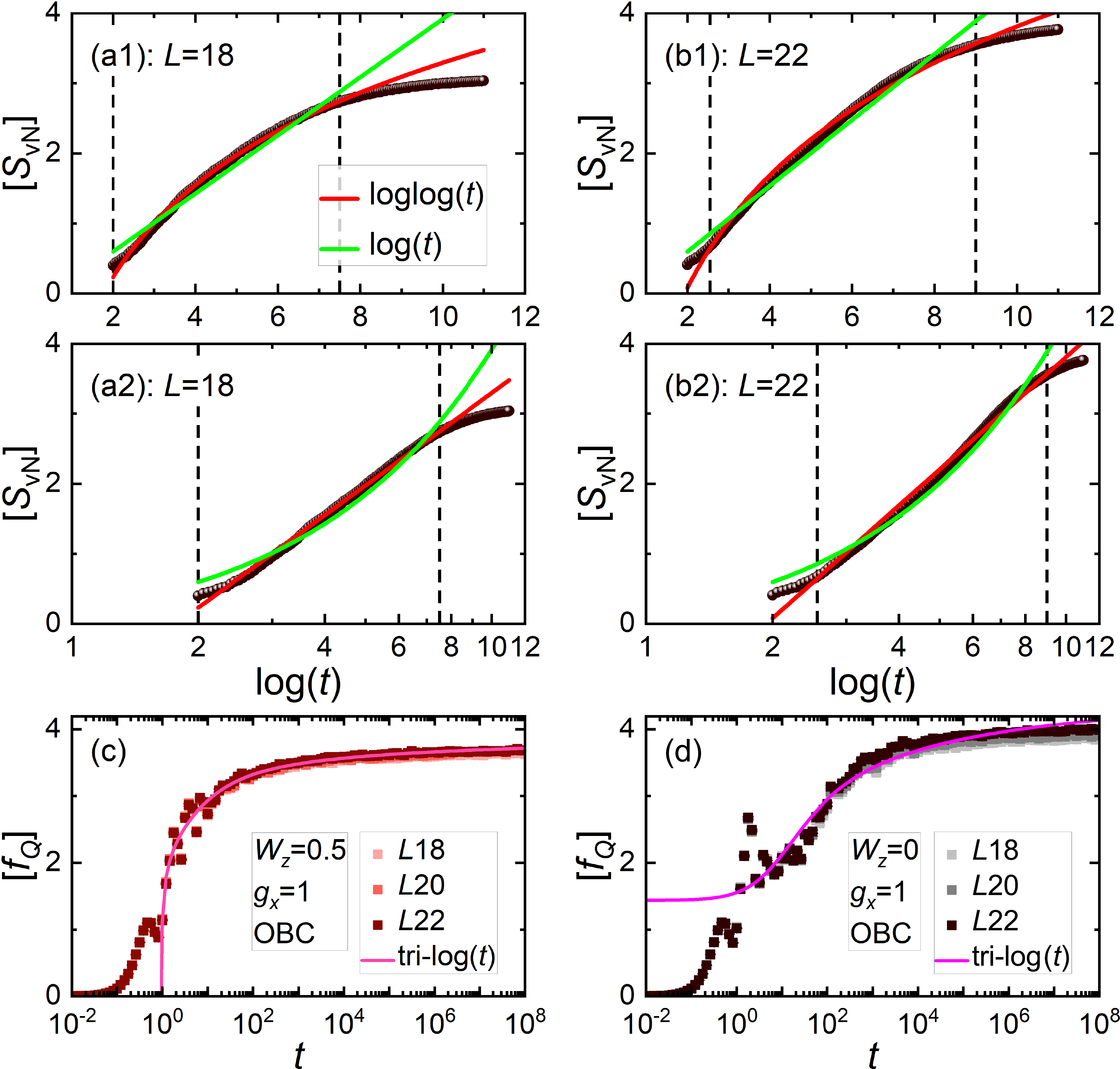}
\caption{\label{fig:fig2b} (a1),(a2): A replot of $\SvN$ in cMBL (stabilized by $\frac{g_x}{W_x}=1,\frac{W_z}{W_x}=0$) on an open chain of $L=18$ but now fitted by two different types of functions: $\log\log(t)$ and $\log(t)$. The lower panel is the same as the upper one but in a semi-log format. Apparently, $\log\log(t)$ gives the better fit. (b1),(b2): The $[S_\textrm{vN}]$ data of a longer open chain $L=22$ with the same model parameters. In each panel, the time interval that matches the double-log fit is marked by two dashed lines. Noticeably, this time window broadens from $t\in(100,10^{7.5})$ for $L=18$ to $t\in(360,10^9)$ for $L=22$. (c),(d): Temporal growth of $[f_Q]$ at two representative points inside the cMBL dome using longer chains. Both can be captured by the triple-log functions of time. (c) corresponds to $W_z=0.5$, while (d) repeats Fig.~\ref{fig:fig2a}(d) for comparison.}
\end{figure}

Experimentally, the QFI, which sets the lower bound of entanglement, was measured in trapped-ion chain \cite{Smith} to witness entanglement growth under the interplay between MBL and long-range interactions. Following \cite{Smith}, we start from N\'eel states in even chains, $|\psi(t\!=\!0)\rangle\!=\!|\!\!\downarrow\uparrow\!\ldots\!\downarrow\uparrow\rangle$, characterized by a staggered $\mathbb{Z}_2$ spin-imbalance operator, ${\I}\coloneqq\frac{1}{L}\sum^L_{i=1}(-1)^i\sigma^z_i$, then the associated QFI density reduces to the connected correlation function of ${\I}$, $f_Q(t)=4L(\langle\psi(t)|{\I}^2|\psi(t)\rangle-\langle\psi(t)|{\I}|\psi(t)\rangle^2)$, which links multipartite entanglement to the fluctuations encoded in measurable quantum correlators. Figure~\ref{fig:fig2a}(b) is a semi-log plot of averaged $[f_Q]$ along the line $g_x=1$ with different $W_z$ color-coded the same way as in Fig.~\ref{fig:fig2a}(a). Likewise, the notable change in functional form of $[f_Q]$ echoes the same cMBL-dMBL eigenstate transition. Specifically, Fig.~\ref{fig:fig2a}(d) shows that the long-time growth of $[f_Q]$ in cMBL matches a triple-log form, which reinforces the double-log fit of $\SvN$ in (c). Parallel relation between $\SvN$ and $[f_Q]$ carries over to the dMBL phase where the power-law growth of $\SvN$ in (e) transforms into a logarithmic growth of $[f_Q]$ in (f).

Table~\ref{table1} recaps the cMBL-dMBL distinction in the fundamental dynamical aspects of entanglement and its witness.

To supplement the ED simulation in main panels, we employ TEBD and matrix-product-operator techniques to verify the cMBL-dMBL transition in larger system sizes $L=28$. A $4$th-order Suzuki-Trotter decomposition is implemented at a maximal time step of unity (the appropriate time step is set by the inverse mean gap and for large $W_z$, we find that smaller time step of about $0.1$ is needed, but as the entanglement growth deep inside dMBL is very slow, TEBD maintains its efficiency at this small time step), and the truncation error per step is controlled lower than $10^{-6}$ by selecting a large bond-dimension threshold $1500$. We check that for $t\lessapprox10^3$ the accumulated total truncation error in typical quasirandom samples is well below $10^{-5}$. The corresponding results and the fits are consistently presented in the insets of Fig.~\ref{fig:fig2a}. However, due to entanglement accumulation, matrix-product-state algorithms of this type retain effectiveness within limited time scales $(t\lessapprox10^3)$.  

\def\arraystretch{1.0}
\setlength{\tabcolsep}{14.5pt}
\begin{table}[b] 
\caption{Hierarchies of dynamic characteristics encompassing constrained, unconstrained, and diagonal MBL phases.} 
\label{table1} 
\centering
\begin{tabular}{ ccc }
 \hline\hline
 \\[-0.9em]
 & $[S_\textrm{vN}]$ & $[$Quantum Fisher Info.$]$ \\ [0.1em] \cline{1-3} \\ [-0.9em]
 cMBL & $\log\log\left(t\right)$ & $\log\log\log\left(t\right)$  \\ [0.1em] \cline{2-3} \\ [-0.9em]
 uMBL & $\log\left(t\right)$ & $\log\log\left(t\right)$  \\ [0.1em] \cline{2-3} \\ [-0.9em]
 dMBL & $t^{\alpha}$ & $\log\left(t\right)$ \\[0.1em]
 \hline\hline
\end{tabular}
\end{table}

{\it Double-log entanglement growth.}---As refinement, Figs.~\ref{fig:fig2b}(a1),(a2) manifest that for cMBL, the double-log fitting function $\log\log(t)$ matches the $\SvN$ data significantly better than the single-log fitting function $\log(t)$, the hallmark of the unconstrained MBL \cite{Znidaric,BardarsonPollmannMoore,Serbyn2013,Luitz2016}. Likewise, Figs.~\ref{fig:fig2b}(b1),(b2) show the entanglement data for larger chain's length $L=22$ with the model parameters, $\frac{g_x}{W_x}=1,\frac{W_z}{W_x}=0$, intact. As indicated by pairs of dashed lines there, it is estimated that the duration that best traces the double-log fit increases progressively from $t\in(100,10^{7.5})\ [\Delta t\approx3.162\times10^7]$ in system $L=18$ to $t\in(360,10^{9})\ [\Delta t\approx10^9]$ in system $L=22$, thus supporting the double-log entanglement buildup in cMBL.

{\it Triple-log rise of QFI.}---Complementarily, Figs.~\ref{fig:fig2b}(c),(d) reproduce the temporal evolutions of $[f_Q]$ for two chosen parameters $W_z=0,0.5$ on longer chains. The fits on top of data from various lengths corroborate the speculation that this triple-log growth of $[f_Q]$ may comprise a feature shared within the dome of cMBL. Similar to Figs.~\ref{fig:fig2b}(a),(b), here we resort to ED and raise the maximum open-chain size to $L=22$.  

\vbox{} \noindent \textbf{Eigenstate transition from transport}

\begin{figure}[tb]
\centering
\includegraphics[width=1\linewidth]{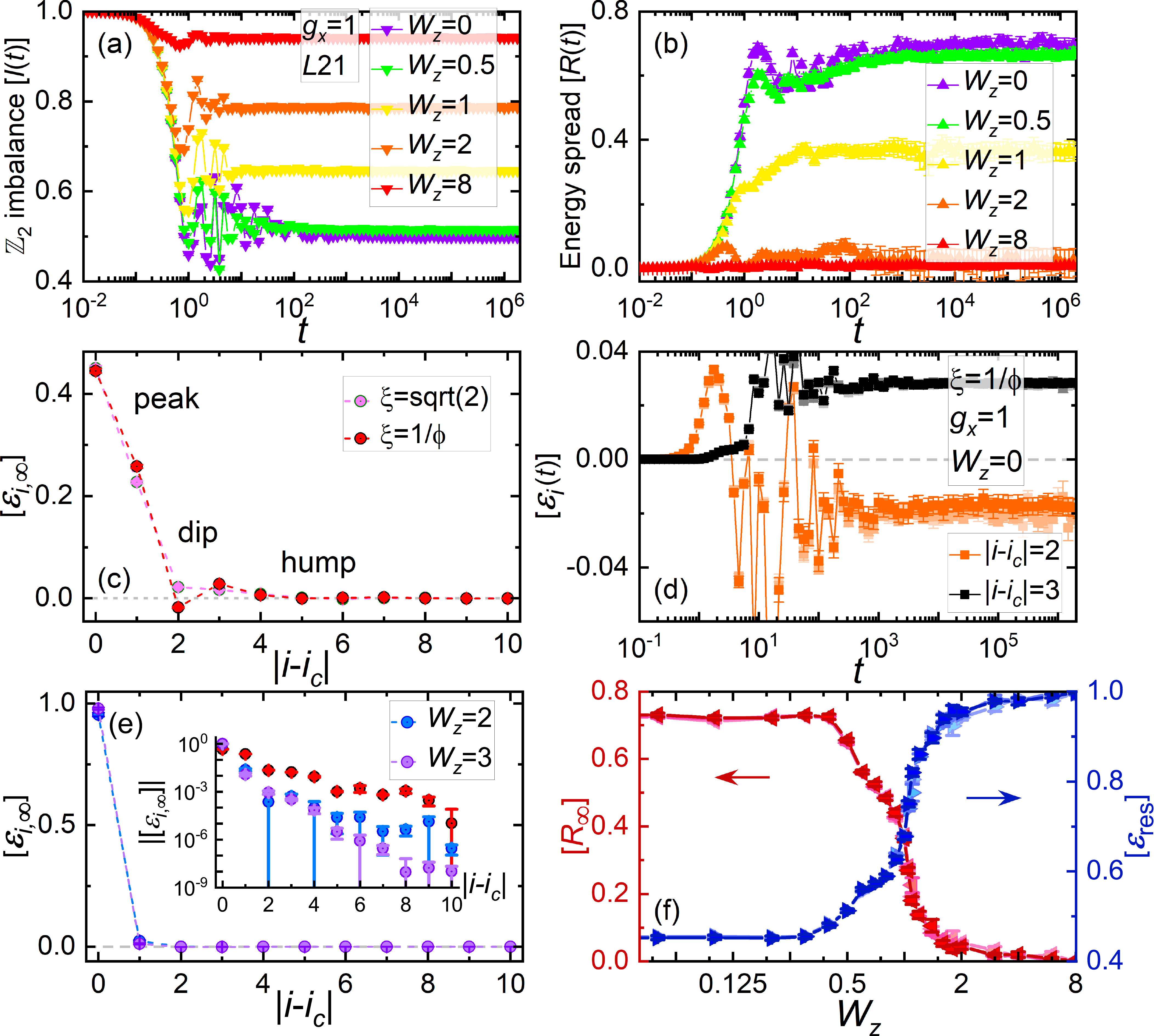}
\caption{\label{fig:fig3a} cMBL-dMBL transition in transport with PBCs and fixed $g_x=1,L=21$. (a),(b): Time evolution of the $\mathbb{Z}_2$ antiferromagnetic imbalance $[{\I}(t)]$ and the energy spread $[{\R}(t)]$ as a function of $W_z$. (c) exemplifies the peak-dip-hump lineshape of $[\varepsilon_{i,\infty}]$ in cMBL for two irrational wavenumbers $\xi$. One set of time-profiles of $[\varepsilon_{i_c\pm2,3}(t)]$ that characterizes the nonmonotonicity of the dip-hump structure is given by (d). (e) shows the lineshape of $[\varepsilon_{i,\infty}]$ in dMBL; the exponential decay can be seen from the semi-log inset wherein the cMBL data (pink dots) are overlaid for comparison. (f): The changes in dynamic \lq\lq order parameters'' $[R_\infty]$ and $[\varepsilon_{\mathrm{res}}]$ under the tuning of $W_z$ signal the transition between cMBL and dMBL. Light to solid colours in (d),(f) correspond to $L=17,19,21$.} 
\end{figure}

\noindent Additionally, there are marked differences between cMBL and dMBL, as reflected through the chain's relaxation from the prepared N\'eel state and the spread of initialized local energy inhomogeneity. In accordance with the time evolution of $\SvN$ and $[f_Q]$, the decay of ${\I}(t)\coloneqq\langle\psi(t)|{\I}|\psi(t)\rangle$ is examined in Fig.~\ref{fig:fig3a}(a). Apart from a quick suppression during $t\lessapprox1$, both MBLs relax to a steady state with finite magnetization. They thus retain remnants of the initial spin configuration in contrast to the thermal phase where $[{\I}(t)]$ vanishes irrevocably. Notice that under the increase of $W_z$, the frozen moment $[{\I}_\infty]$ at infinite $t$ develops monotonously from $\sim0.5$ in cMBL up to $\sim0.9$ in dMBL; before equilibration, the intermediate oscillation of $[{\I}(t)]$ is also damped more severely in dMBL than in (off-diagonal) cMBL.

Following \cite{KimHuse}, the energy transport of the constrained model is scrutinized by monitoring the spread of a local energy inhomogeneity initialized on the central site of an odd chain at infinite temperature, i.e., the system's initial density matrix assumes $\rho(t\!=\!0)=\frac{1}{\dim\!\mathcal{H}}(\mathbb{1}+\varepsilon\X_{\frac{L+1}{2}})$, where $\dim\!\mathcal{H}$ the dimension of projected Hilbert space and $\varepsilon$ the disturbance of energy on site $i_c\coloneqq(L+1)/2$. The quantity measuring the effective distance $\varepsilon$ travels is ${\R}(t)\coloneqq\frac{1}{{\sf Tr}\left[\trho(t)H_\qp\right]}\sum^L_{i=1}\left\{\left|i-i_c\right|{\sf Tr}\left[\trho(t)H_i\right]\right\}$, where $H_i\coloneqq g_i\X_i+h_i\Z_i$ and the time-independent background is subtracted via inserting $\trho(t\!=\!0)\!\coloneqq\!\frac{1}{\dim\!\mathcal{H}}\varepsilon\X_{\frac{L+1}{2}}$. As per ETH, $\varepsilon$ is eventually smeared uniformly over the chain by unitary time evolution and in that circumstance $[{\R}(t\!=\!\infty)]\approx L/4$. Figure~\ref{fig:fig3a}(b) contrasts the behaviour of $[{\R}(t)]$ between cMBL and dMBL. Concretely, for dMBL, $[{\R}]$ stays vanishingly small, thereby $\varepsilon$ remains confined to $i_c$ and shows no diffusion toward infinite $t$. In comparison, as the consequence of a fast expansion within $t\lessapprox100$, largely due to contributions from nearest and next-nearest neighbours, $\varepsilon$ spreads over a finite range of the chain in cMBL. Here, however, the saturated value $[{\R}_{\infty}]$ after an oscillatory relaxation remains subthermal.

\vbox{} \noindent \textbf{Integrals of motion and dynamical order parameters}

\noindent Key distinction between cMBL and dMBL can be further resolved from studying the long-time limit of the spatial distribution of the energy-inhomogeneity propagation. We utilize three quantities to access this information complementarily.

(i) For each quasirandom realization, we parse the definition of $R(t)$ as per the site index, $\varepsilon_i(t)\coloneqq\frac{{\sf Tr}\left[\trho(t)H_i\right]}{{\sf Tr}\left[\trho(t)H_\qp\right]}$, which measures in percentage the extra energy on position $i$ with respect to the total conserved perturbation $\varepsilon$. Observing that $\varepsilon_i$ approaches a constant $\varepsilon_{i,\infty}$ at infinite $t$, one might implement the trick \cite{Anushya}, $\lim\limits_{T\rightarrow\infty}\frac{1}{T}\int^T_0\!O(t)dt\approx\sum\limits_{n}\langle n|O|n\rangle|n\rangle\langle n|$, to extract its value with the aid of randomness,
\beq
\varepsilon_{i,\infty}\coloneqq\varepsilon_i(t\!\rightarrow\!\infty)\approx\frac{\sum\limits_{n}\langle n|\X_{\frac{L+1}{2}}|n\rangle\langle n|H_i|n\rangle}{\sum\limits_{n}E_n\langle n|\X_{\frac{L+1}{2}}|n\rangle},
\label{iom}
\eeq
where $\{|n\rangle\}$ comprises an eigenbasis satisfying $H_\qp|n\rangle=E_n|n\rangle$. Evidently, the profile of $\{\varepsilon_{i,\infty}\}$ bears information pertaining to the local structure of integrals of motion (IOMs).

(ii) The summation of $\varepsilon_{i,\infty}$ weighted by the separation returns the equilibrated value of the effective traveling distance, $R_{\infty}=\sum_{i=1}^L(|i-i_c|\cdot\varepsilon_{i,\infty})$.

(iii) Viewing that the contribution from $i_c$, i.e., the return probability, is missing from $R_{\infty}$, one can define $\varepsilon_{i_c,\infty}$ as the residual energy density on the release place, $\varepsilon_{\mathrm{res}}\coloneqq\varepsilon_{\frac{L+1}{2},\infty}$.

All three quantities defined above are used to distinguish ETH and MBL. Here we show that they are also the dynamical \lq\lq order parameters'' to differentiate between the cMBL and dMBL regimes and identify the transition point therein.

\vbox{} \noindent \textbf{LIOMs and positive definiteness of dMBL}

\noindent Despite the central status of LIOMs in disorder-induced MBL \cite{Ros,ImbrieRosScardicchio,Serbyn,HuseLIOM}, LIOMs in unconstrained aperiodic MBL systems receive attention only recently \cite{Singh,Thomson}. Ref.~\cite{Singh} constructed LIOMs of MBL as time-averaged local operators for interacting fermions subject to aperiodic potentials. They found that in this circumstance $\ell$-bits remain localized even at the vicinity of the quasiperiodic MBL transition. Likewise, upon continuous unitary transforms, Ref.~\cite{Thomson} computed the real-space support of LIOM in quasirandomness-induced MBL and revealed that the effective interactions between LIOMs exhibit features inherited from the underlying aperiodic potential. Interestingly, both works pointed to the weaker finite-size effects in aperiodic modulations than in truly disordered arrangements. Exploiting the instability of LIOMs, they also found that the associated MBL transition may occur at a higher critical quasirandom strength than previously estimated \cite{Khemani}.

Before proceeding to numerics, let's gain some understanding on dMBL within the LIOM framework. The first step forward is to introduce $\breve{Z}_i\coloneqq\mathcal{P}_{i+1}\widetilde{Z}_i\mathcal{P}_{i-1}$ where $\mathcal{P}_i\coloneqq\frac{1}{2}(\mathbb{1}+\sigma^z_i)$ as the building blocks of constrained $\ell$-bits. The convenience of $\breve{Z}_i$ stems from ${\sf Tr}\breve{Z}_i=0$, which contrasts to ${\sf Tr}\widetilde{Z}_i>0$, thereby $\breve{Z}_i$ behaves like a spin free of restrictions. Following \cite{Chen}, it can be proved that as long as $W_z\gg g_x+W_x$, the set of tensor-product operators $\mathcal{I}_L\coloneqq\{\mathcal{Z}_{i_1}\otimes\cdots\otimes\mathcal{Z}_{i_k}\}$ fulfilling $1\leqslant i_1\leqslant i_2\leqslant\cdots i_k\leqslant L,\ i_{a+1}\neq i_a,\ 1\leqslant k\leqslant\frac{L+1}{2}$ may be constructed as a complete, mutually commuting, and linearly-independent basis to express any nontrivial operators that commute with $H_{\textrm{qp}}$; in terms of quasilocal unitaries, $\mathcal{Z}_{i_a}\approx U\breve{Z}_{i_a}U^{\dagger}$. This is because the set of states $\{|\mathcal{Z}_{i_1}\mathcal{Z}_{i_2}\cdots\mathcal{Z}_{i_k}\rangle\}$ derived from $\mathcal{I}_L$ reproduces faithfully the effective eigenbasis of projected Hilbert space for dMBL. Accordingly, the IOM in Eq.~(\ref{iom}) is recast into 
\beq
\mbox{dMBL:}\ \ \sum\limits_{n}\langle n|\widetilde{X}_i|n\rangle|n\rangle\langle n|\approx\sum^{\frac{L-1}{2}}_{m=0}\sum_{r}V^{[i]}_{r,m}\widehat{\mathcal{O}}^{[i]}_{r,m},
\label{expansion_dmbl}
\eeq
where $\widehat{\mathcal{O}}^{[i]}_{r,m}$ denotes the element of $\mathcal{I}_L$ possessing the support on site $i$ (i.e., contains $\mathcal{Z}_i$) and whose furthest boundary from $i$ is of distance $m$. The nonidentical members comprising this specified subset are labelled by $r$. Besides the finite support of $\mathcal{Z}_i$, the key property that promotes $\sum_{n}\langle n|\widetilde{X}_i|n\rangle|n\rangle\langle n|$ to the LIOM of dMBL is the locality condition of its real coefficients, i.e., $V^{[i]}_{r,m} \sim e^{-m/\zeta}$. In terms of LIOM representation, the universal Hamiltonian governing the dynamics of dMBL may assume the following form, $H^{\textrm{dMBL}}_{\textrm{qp}}=\sum_i \widetilde{h}_i\mathcal{Z}_i+\sum_k\sum_{i_1\ldots i_k}J_{i_1\ldots i_k}\mathcal{Z}_{i_1}\mathcal{Z}_{i_2}\cdots\mathcal{Z}_{i_k}$, where from Figs.~\ref{fig:fig2a}(e),(f), it is feasible to infer $J_{i_1\ldots i_k}\sim|i_k-i_1|^{-1/\alpha}\cdot\phi^{-|i_k-i_1|}$, which decays as an exponentially-suppressed power law of LIOMs' separation. Here, $\alpha$ is the same exponent in Table~\ref{table1} and $\phi$ is the golden ratio.

Being the trace of product of two IOMs, one consequence of Eq.~(\ref{expansion_dmbl}) is the positive definiteness of the averaged $[\varepsilon_{i,\infty}]$ featured by an exponential decay in space. Figure~\ref{fig:fig3a}(e) illustrates that this is the case even when $W_z\approx g_x+W_x$.

\vbox{} \noindent \textbf{Peak, dip, hump in cMBL}

\noindent Now we are in the position to highlight the peak-dip-hump structure and the occurrence of negativity in $[\varepsilon_{i,\infty}]$ [see Figs.~\ref{fig:fig3a}(c),(d)] as the peculiarities of cMBL that distinguish it from both dMBL and unconstrained MBL (uMBL) by the presence of pronounced nonlocal correlations. The unambiguous negativity of $[\varepsilon_{i_c\pm2}]$ in Fig.~\ref{fig:fig3a}(c) and the nonmonotonicity of $[\varepsilon_{i_c\pm2,3}]$ in Fig.~\ref{fig:fig3a}(d) point to the insufficiency of Eq.~(\ref{expansion_dmbl}) when addressing the cMBL phase from the dMBL side. Especially, they highlight the dynamical consequence that in cMBL the correlation between the centre site (where the initial energy inhomogeneity locates) and its third nearest neighbours might be stronger than that for its second nearest neighbours, because phenomenologically the net energy current flowing into the second nearest neighbouring sites could appear noticeably less than that flows out. To remedy the inconsistency, we propose as a scenario that the missing pieces may come from the terms in $\mathcal{I}_L$ that are nonlocal with respect to $i$, viz., their support on $i$ vanishes, hence, for cMBL, $\sum_{n}\langle n|\widetilde{X}_i|n\rangle|n\rangle\langle n|\approx\sum^{\frac{L-1}{2}}_{m=0}\sum_{r,\widebar{r}}(V^{[i]}_{r,m}\widehat{\mathcal{O}}^{[i]}_{r,m}+V^{\widebar{[i]}}_{\widebar{r},m}\widehat{\mathcal{O}}^{\widebar{[i]}}_{\widebar{r},m})$. The superscript ${\widebar{[i]}}$ signifies the absence of $\mathcal{Z}_i$ in the associated expansion. Under the successive decrease of $W_z$, it is anticipated that the weights $V^{\widebar{[i]}}_{\widebar{r},m}$ of small $m$ grow significantly such that a finite-size core centred at $i$ forms wherein nonlocal correlated contributions, albeit confined, become predominant. On the contrary, for those $m$ beyond the core, the importance of $V^{\widebar{[i]}}_{\widebar{r},m}$ diminishes sharply so that the rapid decay tail and the overall signatures of localization are well maintained.

Alternatively, the core formation may be monitored by $[R_{\infty}]$ and $[\varepsilon_{\textrm{res}}]$. Figure~\ref{fig:fig3a}(f) illustrates that the duo constitutes the desired \lq\lq order parameters'' from quantum dynamics that take values zero and unity in dMBL and saturate to nontrivial plateaus in cMBL. The critical $W_z$ of the transition is hence estimated to be $\sim0.55$ at $g_x=1$. Furthermore, from Fig.~\ref{fig:fig3a}(c), the core where substantial nonlocal effects take place spans roughly $5$ to $7$ lattice sites which, as per the saturated value of $[R_{\infty}]$ in Fig.~\ref{fig:fig3a}(f), is comparable to a thermal segment of approximately $3$ lattice-spacing long.

This embedded thermal-like core in IOMs plays a crucial role in yielding the novel Lieb-Robinson bound for cMBL. More relevant mathematical justifications are in Ref.~\cite{ChenChenLR}.

\vbox{} \noindent \textbf{cMBL-dMBL transition and return probability}

\begin{figure}[tb]
\centering
\includegraphics[width=1\linewidth]{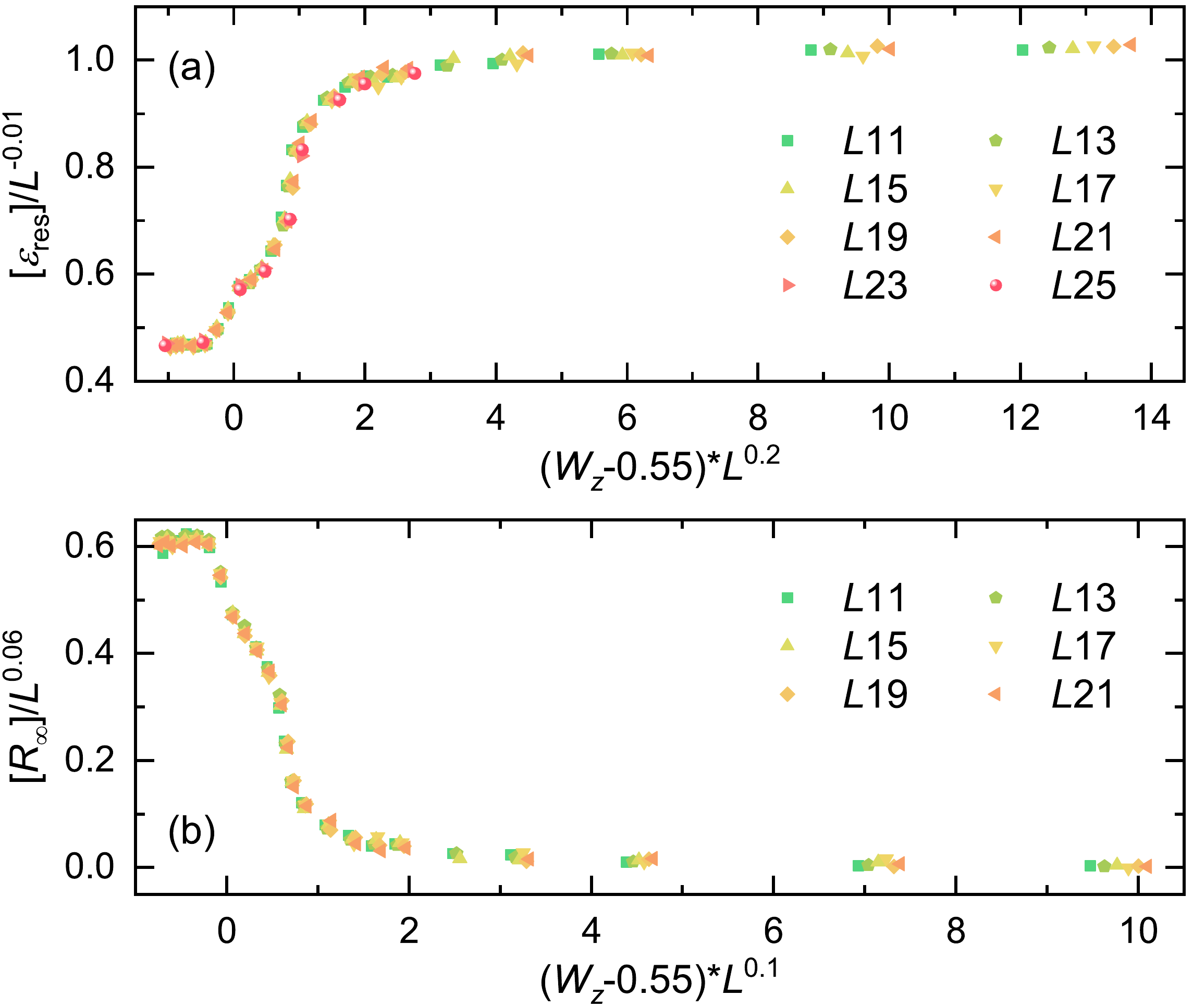}
\caption{\label{fig:fig3b} Finite-size scaling analysis for cMBL-dMBL transition at fixed $g_x=1$. The respective data collapse of random-averaged dynamic order parameters $[\varepsilon_\textrm{res}]$ [panel (a)] and $[R_\infty]$ [panel (b)] yields consistently a critical $W^c_z\sim0.55W_x$. Here, PBCs are used; for $L=23,25$, the results are extracted from the Krylov-typicality approach; other sizes are solved by ED.} 
\end{figure}

\begin{figure}[tb]
\centering
\includegraphics[width=1\linewidth]{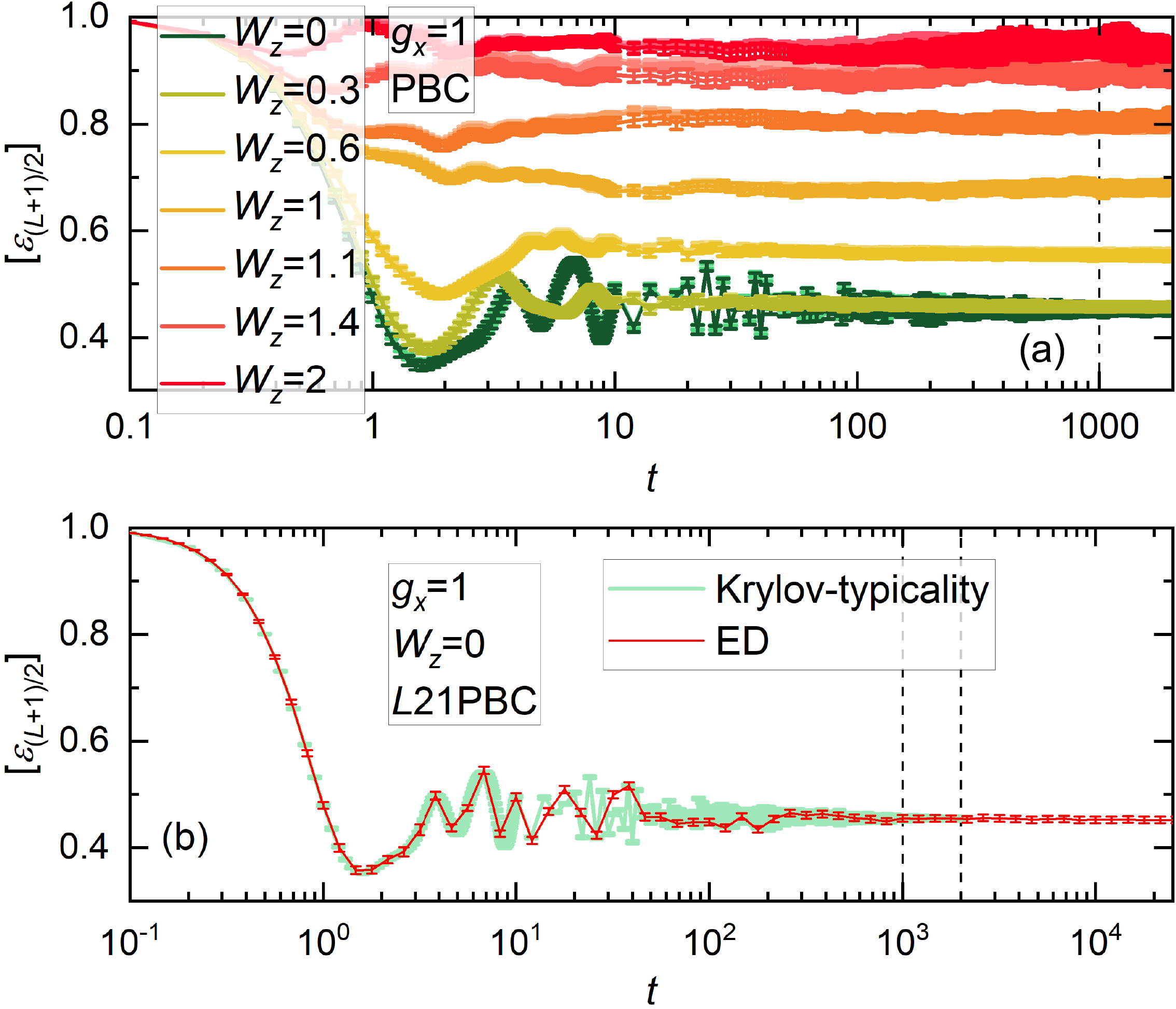}
\caption{\label{fig:fig3c} (a): Time evolution of return probability as a function of $W_z$ at fixed $g_x=1$ obtained by applying Krylov-typicality method to periodic chains. The light to solid colours give the results of $L=21,23,25$. (b): A sample calculation on small chain illustrates that Krylov-typicality approximation reproduces the results of ED.} 
\end{figure}

\noindent In view of the importance of core formation in driving the change between cMBL and dMBL, we perform a finite-size scaling analysis on \lq\lq dynamic order parameters'' $[\varepsilon_\textrm{res}]$ and $[R_\infty]$ to locate more precisely the critical value of $W_z$ that triggers this eigenstate transition. Following Ref.~\cite{Kjall}, the target quantity $Q=[\varepsilon_\textrm{res}]$ (or $[R_\infty]$) at fixed $g_x$ might assume a standard scaling form as follows, $\frac{Q(L,W_z)}{L^{\sigma_Q}}\approx f_Q\!\left((W_z-W^c_{z,Q})L^{\alpha_Q}\right)$, where $f_Q$ is some unknown function associated to $Q$ and similarly $\sigma_Q,\alpha_Q$ are the corresponding scaling exponents. The critical strength $W^c_{z,Q}$ of the transition can then be extracted from the proper data collapse of $Q$.

Figure~\ref{fig:fig3b} depicts scaling profiles of $[\varepsilon_\textrm{res}]$ and $[R_\infty]$ at fixed $g_x=1$ for different system sizes ranging from $L=11$ to $25$ using PBCs. As shown by panel (a), the scaling collapse of the data for $[\varepsilon_\textrm{res}]$ yields the following set of parameters, $W^c_{z,\varepsilon}\approx0.55,\ \sigma_\varepsilon\approx-0.01,\ \alpha_\varepsilon\approx0.2$.

Although $[\varepsilon_\textrm{res}]$ and $[R_\infty]$ carry compatible information, as $[\varepsilon_{i_c}(t)]$ stands for return probability, an autocorrelation function easier to measure in experiments, we extend its evaluation to longer chains $L=23,25$ by resort to Krylov-typicality technique (see methods section). Figure~\ref{fig:fig3c}(a) presents the obtained $[\varepsilon_{\frac{L+1}{2}}(t)]$ for a range of $W_z$ close to the transition of $g_x=1$. It is noticeable that due to localization, all evolution curves relax to their constant lineshapes whose saturation values strengthen with $W_z$ and form two individual plateaus around $\sim0.45$ for cMBL and $\sim1$ for dMBL, respectively. This convergent trend allows for an estimate of $[\varepsilon_\textrm{res}]$ by averaging the return probability over a later period $t\in[1000,2000]$, which produces the data points of $L=23,25$ in Fig.~\ref{fig:fig3b}(a). The reliability of such a procedure is justified by Fig.~\ref{fig:fig3c}(b) where we check the correctness of Krylov-typicality approximation and the appropriateness of the chosen time window via a benchmark test against the exact results of $L=21$.

Because energy diffusion is ceased in localized regions (in other words, the initial energy imbalance is restricted to the chain centre), adding more sites on the chain ends generates small finite-size flows of $[\varepsilon_\textrm{res}]$ seen in Fig.~\ref{fig:fig3b}(a) even after raising the length limit to $L=25$. Likewise, as plotted by Fig.~\ref{fig:fig3b}(b), the estimates of scaling parameters obtained from the data collapse of $[R_\infty]$ read $W^c_{z,R}\approx0.55,\ \sigma_R\approx0.06,\ \alpha_R\approx0.1$.

\vbox{} \noindent \textbf{DISCUSSION}

\noindent To conclude, we find a cMBL regime and a probably discontinuous cMBL-cETH transition in the quasirandom Rydberg-blockade chain. The orthogonality between the field strength and the projection direction renders cMBL and its discontinuous eigenstate transition fundamentally different from dMBL, uMBL, and the continuous uMBL-uETH transition. Particularly, the entanglement entropy in cMBL grows as a double-log function of time, as opposed to the power-law growth in dMBL and the single-log growth in uMBL.

The presumed discontinuity of the cMBL-cETH transition is evidenced numerically. A future analytical elucidation of its underpinnings may potentially improve the existing theoretical framework for MBL transition in a substantial way.

Even though LIOMs capture the phenomenology of dMBL, the cMBL-dMBL transition triggered by the rotation of the field orientation accentuates the importance of nonlocal components in the IOMs of cMBL, which, together with the double-log entanglement growth, raises doubts about how to define the meaningful LIOMs and the universal Hamiltonian suitable for cMBL.

The continual investigations on these open questions promise to further our understanding of unconventional MBL beyond the current scope.

\vbox{} \noindent \textbf{METHODS}

\noindent \textbf{Quantum dynamics computations}

\noindent In this work, to cope with the many-body nonequilibrium problem subject to intertwining complexities from constraint and randomness, three numerical approaches, ED, Krylov-typicality, and TEBD, are employed. 

{\it ED.}---For small chains, we resort to the ED method to access the long-time limit, where quadruple precision is implemented for achieving the time scale up to $t\approx 10^{29}$. Within full diagonalization, the infinite-time limit is resolvable by invoking the diagonal approximation. Further, rather than removing the $\downarrow\downarrow$-motifs from unconstrained Hilbert space, we construct the projected spin-$\frac{1}{2}$ basis as a selected set of binary numbers by fulfilling the constraint rule using combinatorial reasoning, which is more efficient for larger system sizes.

{\it TEBD.}---One alternative to evaluate the time evolution of longer quantum spin chains, albeit with the limitation of much shorter time scales, is the TEBD algorithm \cite{Vidal}, which is built upon parametrization of a quantum wavefunction in terms of matrix-product states (MPS) \cite{Schollwock},
\beq
|\psi\rangle=\sum_{\sigma_1,\ldots,\sigma_L}A^{[1]}A^{[2]}\cdots A^{[L]}|\sigma_1,\ldots,\sigma_L\rangle,
\label{MPS}
\eeq
where $A^{[i]}$ stands for a three-leg tensor at site $i$ carrying one physical bond $\sigma_i=1,2$ for a local spin-$\frac{1}{2}$ system and two virtual legs of dimension $\chi^i_L$ and $\chi^i_R$. TEBD relies on the low amount of entanglement generation and the Suzuki-Trotter decomposition of the time evolution operator. Concretely, at $4$th order, this unitary can be approximated in a symmetric format \cite{Schollwock},
\beq
e^{-iH\tau}=U(\tau_1)U(\tau_2)U(\tau_3)U(\tau_2)U(\tau_1)+\mathcal{O}(\tau^5),
\eeq
where
\begin{align}
U(\tau_i)&=e^{-iH_{\textrm{odd}}\tau_i/2}e^{-iH_{\textrm{even}}\tau_i}e^{-iH_{\textrm{odd}}\tau_i/2}, \\
\tau_1=\tau_2&=\frac{\tau}{4-\sqrt[3]{4}},\ \ \ \ \ \tau_3=\tau-2(\tau_1+\tau_2),
\end{align}
and we assume that the total inspected Hamiltonian $H$ comprises a sum of two-site operators that can be divided into the respective $H_{\textrm{even}}$ and $H_{\textrm{odd}}$ parts living across the even and odd bonds. Evidently, starting from an arbitrary product state in the projective spin basis, to a good approximation, the repeated application of the unitary time evolution will not generate components that violate the constraint.

The calculation of QFI entails the evaluation of $I^2$, which is easily computed within TEBD via recasting $I$ as a matrix-product operator (MPO) \cite{Schollwock}: analogous to the MPS representation in Eq.~(\ref{MPS}), a generic operator $O$ is rewritten as
\beq
O=\sum_{\substack{\sigma_1,\ldots,\sigma_L \\ \sigma'_1,\ldots,\sigma'_L}}W^{[1]}W^{[2]}\cdots W^{[L]}|\sigma_1,\ldots,\sigma_L\rangle\langle\sigma'_1,\ldots,\sigma'_L|,
\eeq
where $W^{[i]}$ is a four-leg tensor on site $i$ equipped with two physical bonds $\sigma_i,\sigma'_i$ and two virtual bonds of dimension $D\times D$. For spin-imbalance operator $I$, $W$'s are simply given by
\beq
W^{[i]}=\begin{bmatrix}
\mathds{1} & f_i\sigma^z_i \\[0.5em]
0 & \mathds{1}
\end{bmatrix},
\eeq
where $f_i=(-1)^i/L$, and $\mathds{1},\sigma^z_i$ are $2\times2$ Pauli matrices, therefore $D=2$ in this case. Then, $I^2$ consists of a stacking of two identical layers of $W$'s tensors, whose expectation value at $t$ is obtained by executing optimal contractions.

{\it Krylov-typicality.}---For brevity, let us recap the main steps of Krylov-space technique. By definition, starting from an arbitrary normalized wave-vector $|\phi_0\rangle$, the associated linearly-independent Krylov subspace $\mathcal{K}$ is generated by consecutively applying the Hamiltonian $H$ onto $|\phi_0\rangle$ $(m-1)$ times,
\beq
\mathcal{K}\coloneqq\left\{|\phi_0\rangle,H|\phi_0\rangle,H^2|\phi_0\rangle,\cdots,H^{m-1}|\phi_0\rangle\right\}.
\eeq
Through clever recombination, an equivalent but more convenient reformulation of $\mathcal{K}$ exists, which is mutually orthonormal and called the Lanczos basis derived from $|\phi_0\rangle=|v_0\rangle$,
\beq
\mathcal{K}\sim\mathcal{L}\coloneqq\left\{|v_0\rangle,|v_1\rangle,|v_2\rangle,\cdots,|v_{m-1}\rangle\right\},
\eeq
where, to remedy the loss of orthogonality, the procedure of reorthogonality is always assumed. The advantage of $\mathcal{L}$ lies in the fact that for most practical calculations, it suffices to choose the Lanczos dimension $m\approx50$ to $100$, which is orders of magnitude smaller than the full Hilbert-space dimension $\mathcal{D}$.

The above rationale can be recapitulated in terms of the following basis transformation,
\beq
F^\dagger H F=H_{\textrm{lanc}},
\eeq
where the full Hamiltonian $H$ is written in the original physical basis, while the heavily reduced Hamiltonian $H_{\textrm{lanc}}$ is recast in the Lanczos basis specific for the neighbourhood of $|v_0\rangle=|\phi_0\rangle$. It is easy to prove that for Hermitian operator $H$, $H_{\textrm{lanc}}$ is a tridiagonal matrix. Moreover, stacking the Lanczos states yields the fundamental transformation matrices $F,F^\dagger$,
\begin{gather}
F=\left[|v_0\rangle\ |v_1\rangle\ |v_2\rangle\ \cdots\ |v_{m-1}\rangle\right]_{\mathcal{D}\times m}, \\[0.5em]
F^\dagger=\begin{bmatrix}
\langle v_0| \\
\langle v_1| \\
\vdots \\
\langle v_{m-1}|
\end{bmatrix}_{m\times\mathcal{D}}.
\end{gather}
The essence of Lanczos approximation can then be encapsulated in terms of the following single relation,
\beq
FF^\dagger\approx\mathds{1}_{m\times m},
\label{complete}
\eeq
which is exact iff $m$ equals the Hilbert-space dimension $\mathcal{D}$. 

Armed with these preparations, we are ready to derive the formula for the real-time propagation of a normalized vector $|\psi(t)\rangle=|\phi_0\rangle=|v_0\rangle$ under the unitary evolution of the Hamiltonian $H$ up to a small time decimation $\delta$, i.e.,
\begin{align}
|\psi(t+\delta)\rangle\approx F V_{\textrm{lanc}}e^{-i\frac{\delta}{\hbar} D_{\textrm{lanc}}}V^\dagger_{\textrm{lanc}}F^\dagger|\psi(t)\rangle,
\label{psitt}
\end{align}
where the tridiagonal Lanczos matrix $H_{\textrm{lanc}}$ is diagonalized by the unitary matrix $V_{\textrm{lanc}}$, $H_{\textrm{lanc}}=V_{\textrm{lanc}}D_{\textrm{lanc}}V^\dagger_{\textrm{lanc}}$. Symbolically, one writes the combined vector $F^\dagger|\psi(t)\rangle$ in the explicit form,
\beq
F^\dagger|\psi(t)\rangle=\begin{bmatrix}
\frac{\langle \psi(t)|\psi(t)\rangle}{\sqrt{\langle \psi(t)|\psi(t)\rangle}} \\
0 \\
\vdots \\
0
\end{bmatrix}_{m\times1},
\eeq 
where the zeros are resultant from the orthogonality between different Lanczos basis states.

Notice that both $|\psi(t)\rangle$ and $|\psi(t+\delta)\rangle$ are $\mathcal{D}\times1$ vectors in the computational basis, therefore the evaluation of the half-chain von Neumann entanglement entropy at $t+\delta$ is proceeded in the usual way once $|\psi(t+\delta)\rangle$ is available.

For an initial energy inhomogeneity at the central site of a Rydberg chain, its return probability under the unitary time evolution of the quasiperiodic Hamiltonian $H_{\textrm{qp}}$ is
\beq
\varepsilon_{\frac{L+1}{2}}(t)\coloneqq\frac{1}{{\sf Tr}\left[\widetilde{\rho}(t)H_{\textrm{qp}}\right]}{\sf Tr}\left[\widetilde{\rho}(t)H_{\frac{L+1}{2}}\right].
\label{Returneff}
\eeq
Here we assume the infinite temperature. The initial density matrix $\widetilde{\rho}(t\!=\!0)$ specifies the spatial distribution of the energy disturbance at $t=0$, whose subsequent dynamics can be couched in the Heisenberg representation as $(\hbar=1)$,
\begin{gather}
\widetilde{\rho}(t)\coloneqq e^{iH_{\textrm{qp}}t}\cdot\widetilde{\rho}(t=0)\cdot e^{-iH_{\textrm{qp}}t}, \label{rho_t} \\
\widetilde{\rho}(t=0)\coloneqq\frac{\varepsilon}{\mathcal{D}}\widetilde{X}_{\frac{L+1}{2}}. \label{rho_0}
\end{gather}
Note that $H_{\textrm{qp}}$ is discretized in Eq.~(\ref{Returneff}), i.e.,
\begin{align}
H_\textrm{qp}=\sum^L_{i=1}H_i,\ \ \ \ \ \mbox{where}\ \ \ \ \ H_i=g_i\widetilde{X}_i+h_i\widetilde{Z}_i.
\end{align}

There exists one extra complication in the computation of $\varepsilon_{\frac{L+1}{2}}(t)$. By full diagonalization, the trace over the entire Hilbert space is accomplishable through summing up the contributions of all eigenvectors with the equal weight,
\beq
{\sf Tr}[\ldots]=\sum^\mathcal{D}_{n=1}\langle E_n|\ldots|E_n\rangle.
\label{trace}
\eeq
This approach is impractical once $L\geqslant23$. For instance, when $L=23$ under PBCs, $\mathcal{D}=64079$ and the required ram to compute a single random realization is over $190$GB.

To make progress, we invoke the trick of dynamical typicality \cite{Popescu,Goldstein,Steinigeweg} to approximately evaluate the trace of an operator over the gigantic Hilbert space. The key idea is to replace Eq.~(\ref{trace}) by a single scalar product using a pure state $|\Psi_{\textrm{gaus}}\rangle$,
\beq
\frac{1}{\mathcal{D}}{\sf Tr}[\ldots]=\frac{1}{\mathcal{D}}\sum^\mathcal{D}_{n=1}\langle E_n|\ldots|E_n\rangle\approx\langle\Psi_{\textrm{gaus}}|\ldots|\Psi_{\textrm{gaus}}\rangle,
\label{typicality}
\eeq 
where $|\Psi_{\textrm{gaus}}\rangle$ is written in the computational basis with dimension $\mathcal{D}$ whose entries are Gaussian random numbers with zero means. The Gaussian distribution of the complex components of $|\Psi_{\textrm{gaus}}\rangle$ guarantees that $|\Psi_{\textrm{gaus}}\rangle$ is drawn uniformly on the hypersphere of the full Hilbert space (i.e., according to the Haar measure) such that the corresponding probability distribution is invariant under all unitary transformations within the Hilbert space. As per formal theory of typicality \cite{Popescu,Goldstein}, $|\Psi_{\textrm{gaus}}\rangle$ could be an effective representative of the underlying statistical ensemble.

Practically, $|\Psi_{\textrm{gaus}}\rangle$ is constructed in a simple manner,
\beq
|\Psi_{\textrm{gaus}}\rangle=\frac{1}{\mathcal{N}}\sum^{\mathcal{D}}_{a=1}\left(r_a+is_a\right)|a\rangle,
\eeq
where $|a\rangle$ enumerates the physical computational basis, and $r_a,s_a$ are real, independent Gaussian random numbers with mean zero and variance one, $\mathcal{N}=\sqrt{\sum^\mathcal{D}_{a=1}\left[r^2_a+s^2_a\right]}$. As $\mathcal{D}$ is at the order of $10^7$ (for example, $\mathcal{D}=1346269$ for $L=29$ under OBCs), for averages over $2000$ independent quasirandom realizations, a good random number generator with long period $(\sim10^{18})$ might be needed.

In real computation, the accuracy of dynamical typicality can be improved by using multiple pure Gaussian states in a single evaluation of Eq.~(\ref{typicality}). To maximize the overall landscape for the random realizations of $H_\textrm{qp}$ and $|\Psi_{\textrm{gaus}}\rangle$, it is economic to invoke one independent $|\Psi_{\textrm{gaus}}\rangle$ for each different quasiperiodic $H_\textrm{qp}$, and then perform the average over $2000$ such joint samples. Accordingly, under the assumption of the validity of dynamical typicality, the random-averaged return probability might be approximated as follows,
\beq
[\varepsilon_{\frac{L+1}{2}}(t)]\approx\frac{1}{\mathcal{R}}\sum^{\mathcal{R}}_{q=1}\langle\Psi^q_{\textrm{gaus}}|\left[\widetilde{X}^q_{\frac{L+1}{2}}(t)\widetilde{X}_{\frac{L+1}{2}}\right]|\Psi^q_{\textrm{gaus}}\rangle,
\label{Returneff_averaged}
\eeq
where $q$ denotes the involvement of the $q$th random sample and the total number for the joint random samples is over $2000$, $\mathcal{R}\geqslant2000$. Being a scalar product of pure state, each summand in Eq.~(\ref{Returneff_averaged}) is evaluable by the Lanczos method.

For all results presented in this paper, we perform the random sample calculations over at least $1000$ independent quasiperiodic configurations of the model parameters $\phi_x,\phi_z$ and when evaluating entanglement growth, the initial product states are additionally selected from the constrained spin basis in another randomized manner \cite{BardarsonPollmannMoore}. The corresponding statistical uncertainties are estimated from the normal variance of varied averaged quantities as per \cite{Oganesyan}.

\vbox{} \noindent \textbf{DATA AVAILABILITY}

\noindent The data set that supports the findings of the present study can be available from the corresponding authors via email upon reasonable request.

\vbox{} \noindent \textbf{ACKNOWLEDGEMENTS}

\noindent The discussion with M. Heyl was acknowledged. This work is supported by the SKP of China (Grant Nos. 2016YFA0300504 and 2017YFA0304204) and the NSFC Grant No. 11625416.

\vbox{} \noindent \textbf{AUTHOR CONTRIBUTIONS}

\noindent All authors contributed equally to this work.

\vbox{} \noindent \textbf{COMPETING INTERESTS}

\noindent The authors declare no competing financial or non-financial interests.

\bibliography{cMBL}

\end{document}